\documentclass[authoryear,12pt]{article}

\usepackage[margin=1in]{geometry}
\usepackage{setspace}
\usepackage{amsmath}
\usepackage{amsthm}
\usepackage{graphicx}
\usepackage{caption}
\usepackage{subcaption}
\usepackage{natbib}
\usepackage{authblk}
\usepackage{lineno}

\begin{document}

\title{The evolution of queen control over worker reproduction in the social Hymenoptera}
\author{Jason~Olejarz$^\mathrm{a}$, Carl~Veller$^\mathrm{a,b}$, and Martin~A.~Nowak$^\mathrm{a,b,c}$}
\affil{$^\mathrm{a}$Program for Evolutionary Dynamics, Harvard University, Cambridge, MA~02138 USA\\$^\mathrm{b}$Department of Organismic and Evolutionary Biology, Harvard University, Cambridge, MA~02138 USA\\$^\mathrm{c}$Department of Mathematics, Harvard University, Cambridge, MA~02138 USA\\jolejarz@fas.harvard.edu; carlveller@fas.harvard.edu; martin\_nowak@harvard.edu}
\date{}

\maketitle

\begin{abstract}
A trademark of eusocial insect species is reproductive division of labor, in which workers forego their own reproduction while the queen produces almost all offspring. The presence of the queen is key for maintaining social harmony, but the specific role of the queen in the evolution of eusociality remains unclear. A long-discussed scenario is that a queen either behaviorally or chemically sterilizes her workers. However, the demographic and ecological conditions that enable such manipulation are unknown. Accordingly, we propose a simple model of evolutionary dynamics that is based on haplodiploid genetics. We consider a mutation that acts in a queen, causing her to control the reproductive behavior of her workers. Our mathematical analysis yields precise conditions for the evolutionary emergence and stability of queen-induced worker sterility. These conditions do not depend on the queen's mating frequency. Moreover, we find that queen control is always established if it increases colony reproductive efficiency and can evolve even if it decreases colony efficiency. We further outline the conditions under which queen control is evolutionarily stable against invasion by mutant, reproductive workers.
\end{abstract}

\vspace*{10mm}

\section*{Keywords}

\noindent
Chemical Communication, Pheromones, Reproductive Division of Labor, Social Insects, Natural Selection, Evolutionary Dynamics

\vspace*{10mm}

\section{Introduction}
\label{sec:intro}

Many species of ants, bees, and wasps form highly complex eusocial societies characterized by dominance hierarchies and reproductive division of labor \citep{Wilson_1971,Michener_1974,Holldobler_1990,Gadagkar_2001,Hunt_2007}.  In most cases, both the queen and the workers are capable of laying male eggs parthenogenetically, but the workers often forego their own reproduction, allowing the queen to produce the majority of drones \citep{Bourke_1988,Heinze_2004,Ratnieks_2006}.

There are several ways in which this behavior could arise.  One possibility is that a `policing' mutation acts in a worker, causing that worker to destroy male eggs produced by other workers \citep{Ratnieks_1988,Olejarz_2016}.  Alternatively, a `non-reproduction' mutation could act in a worker, causing that worker to forego its own reproduction \citep{Olejarz_2015,Doebeli_2015}.  Such mutations can spread and eventually fix in the population if the resulting gains in colony reproductive efficiency are sufficiently large \citep{Ratnieks_1988,Olejarz_2016,Olejarz_2015}.  Finally, and as the case we consider in this paper, a mutation could act in a queen, causing her to suppress her workers' reproduction \citep{Bourke_1988,Holldobler_1990,Vienne_1998}.

There are several mechanisms by which a queen can manipulate her workers' reproductive output.  In small colonies, the queen or dominant individual can directly control worker reproduction by eating worker-laid eggs or by aggressing workers who attempt to lay eggs \citep{Wilson_1971,Michener_1974,Oster_1978,Heinze_1994,Bourke_1995,Koedam_1997,Dapporto_2010,Smith_2011}. Indirect chemical suppression of worker reproduction is also possible through queen pheromones \citep{Keller_1993,Konrad_2012,Richard_2013,Nunes_2014,Oi_2015a,Leonhardt_2016}, which are especially important in species with large colonies, where direct queen policing is infeasible \citep{Gadagkar_1997,Katzav-Gozansky_2006,LeConte_2008}. 

Pheromonal suppression by queens or dominant individuals has long been recognized in the eusocial Hymenoptera \citep{Keller_1993,Kocher_2011}. For example, queen tergal gland secretions \citep{Wossler_1999} and queen mandibular pheromone \citep{Hoover_2003} have both been shown to limit ovarian development in honey bee workers (genus \textit{Apis}), while in the carpenter ant \textit{Camponotus floridanus}, worker-laid eggs experimentally marked with the queen-derived surface hydrocarbons were significantly less likely to be destroyed by other workers \citep{Endler_2004}.  Pheromonal suppression of worker reproduction has also been documented in primitively eusocial species, including the polistine wasps \textit{Polistes dominula} \citep{Sledge_2001} and \textit{Ropalidia marginata} \citep{Bhadra_2010,Saha_2012,Mitra_2014}, the euglossine bee \textit{Euglossa melanotricha} \citep{Andrade-Silva_2015}, and several species in \textit{Bombus} \citep{Ayasse_2014,Holman_2014}.

Despite the ubiquity of the phenomenon, a rigorous theoretical understanding of the evolution of queen suppression of worker reproduction is lacking.  What are the precise conditions under which queen control evolves?  What demographic and ecological characteristics of insect populations result in the evolutionary emergence of queen control? To address these questions, we formulate a model of population dynamics that is based on haplodiploid genetics \citep{Nowak_2010,Olejarz_2015,Olejarz_2016}. In this model, we study the population genetics of alleles, dominant or recessive, that act in queens to reduce worker reproduction. We derive exact conditions for invasion and stability of these alleles, for any number of matings of the queen, and interpret these conditions in terms of the colony efficiency effects of suppressing worker reproduction.

A related, longstanding debate in the literature concerns the nature of queen chemical suppression of worker reproduction in terms of workers' `evolutionary interests' \citep{Keller_1993,LeConte_2008,Heinze_2009}. Should queen chemical suppression be interpreted as coercive control of workers (against their evolutionary interests), or are these chemicals best thought of as honest signals of queen presence or fertility (so that their induction of non-reproduction in workers can in fact be in those workers' evolutionary interests)?  Empirical studies provide support for both interpretations \citep{Keller_1993,Katzav-Gozansky_2006,LeConte_2008,Strauss_2008,Kocher_2009,Heinze_2009,Holman_2010,vanZweden_2010,Maisonnasse_2010,Brunner_2011,Kocher_2011,Peso_2015}. 

Our setup, based on population genetics, offers a simple and attractive framework for classifying queen suppressor chemicals as either coercive or honest signals. Suppose a queen suppressor mutation has fixed, so that all queens produce chemicals that suppress workers' reproduction. Now suppose that a `resistance' mutation arises that renders workers in whom it is expressed immune to queen suppressor chemicals, so that these workers again lay male eggs. If this `resistance' mutation invades, then resistance is seen to be in the workers' evolutionary interests, and the initial queen suppression should be interpreted as coercive. If not, then we interpret the queen suppressor chemical to be an honest signal \citep{Gonzalez_2013}. Invadability of the population by this rare `resistance' allele is equivalent to evolutionary instability of a non-reproduction allele acting in workers, the formal population genetical conditions for which are given in \cite{Olejarz_2015}. We use these conditions to distinguish the demographic and ecological parameter regimes in which queen suppression should be thought of as coercion or as honest signalling. We also explore the similarly relevant possibility of partial queen control inducing complete worker sterility \citep{Bourke_1988,Ratnieks_2006}.

\vspace*{10mm}

\section{Model}
\label{sec:model}

We study queen control of workers in the context of haplodiploid sex determination, as found in ants, bees, and wasps. Fertilized eggs (diploid) become females, and unfertilized eggs (haploid) become males. 

\begin{figure}
\centering
\centering
\includegraphics*[width=1\textwidth]{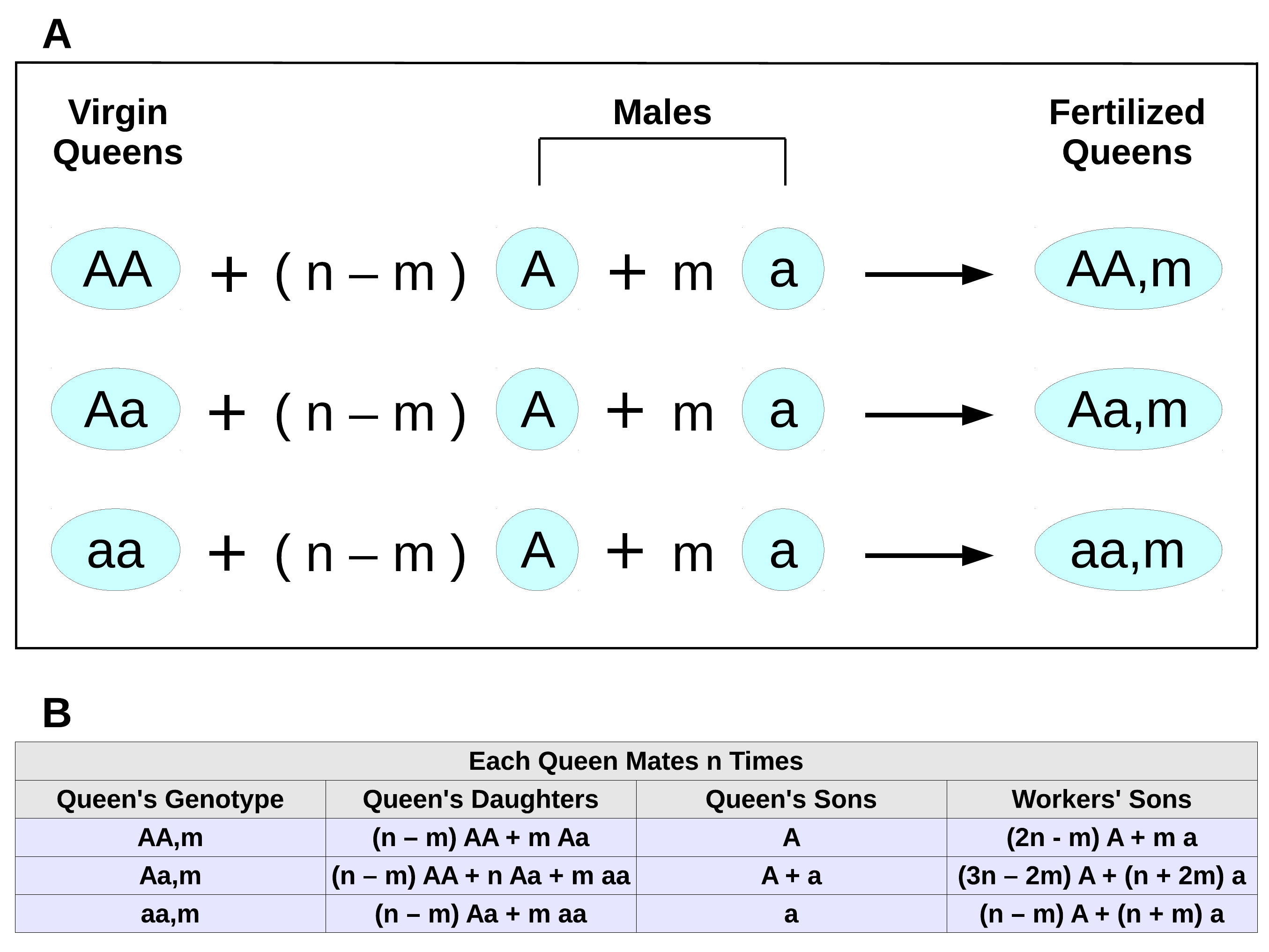}
\caption{The mating events are shown in (a).  The reproduction events are shown in (b).}
\label{fig:colonies}
\end{figure}

A single gyne mates with $n$ distinct, randomly-chosen drones. She then founds a colony and becomes its queen (Figure \ref{fig:colonies}(a)). She fertilizes haploid eggs with the sperm from each of the $n$ males that she mated with to produce diploid female eggs. When these female eggs hatch, the resulting individuals become workers in the colony. In addition, the queen produces unfertilized haploid male eggs. Workers can also produce haploid male eggs, leading to reproductive conflict over male production within a colony (Figure \ref{fig:colonies}(b)).

We consider the evolutionary dynamics of two alleles---a wild-type allele, $A$, and a mutant allele, $a$. We use the following notation for individuals of various genotypes. There are two types of drones: $A$ and $a$. There are three types of gynes: $AA$, $Aa$, and $aa$. A queen's type (or, equivalently, that of a colony, since each colony is headed by a single queen) is denoted by $AA,m$; $Aa,m$; or $aa,m$, depending on whether the queen's own genotype is $AA$, $Aa$, or $aa$, respectively, and the number, $m$, of mutant (type $a$) drones she mated with, requiring $0 \leq m \leq n$.  We use the notation $X_{AA,m}$, $X_{Aa,m}$, and $X_{aa,m}$ to denote the frequencies of the colony types in the population, and we require that $\sum_{m=0}^n (X_{AA,m}+X_{Aa,m}+X_{aa,m})=1$ at all times.

The mutant allele, $a$, acts in a queen to alter her phenotype. If the mutant allele, $a$, is dominant, then type $AA,m$ queens are wild-type, while type $Aa,m$ and type $aa,m$ queens have the mutant phenotype. If the mutant allele, $a$, is recessive, then type $AA,m$ and type $Aa,m$ queens are wild-type, while type $aa,m$ queens have the mutant phenotype.

In colonies headed by wild-type queens, a fraction $0 \leq p \leq 1$ of males are produced by the queen, and new gynes and drones are produced at rate $r \geq 0$.  In colonies headed by queens with the mutant phenotype, a fraction $0 \leq p' \leq 1$ of males are produced by the queen, and new gynes and drones are produced at rate $r' \geq 0$.  Thus, colonies headed by queens with the mutant phenotype have different values of the fraction of queen-produced males and colony efficiency---$p'$ and $r'$, respectively---compared with colonies headed by wild-type queens.

Importantly, our mathematical analysis is robust.  It does not make restrictive or nuanced assumptions about the underlying factors that influence the values of $p$, $r$, $p'$, and $r'$ in any particular case of interest.  The values of the parameters $p$, $r$, $p'$, and $r'$ indeed result from interplay between many demographic and ecological factors.  It is instructive to consider the relative values of these parameters in the context of a queen that influences her workers' reproduction.  We expect that $p'>p$; i.e., the effect of the queen's manipulation is to increase the fraction of male eggs that come from her. $r'$ may be greater than or less than $r$. If $r'>r$, then the queen's manipulation effects an increase in colony efficiency, while if $r'<r$, then the queen's manipulation effects a decrease in colony efficiency.

\vspace*{10mm}

\section{Results}
\label{sec:results}

The key question is:  What values of the parameters $p$, $r$, $p'$, and $r'$ support the evolution of queen suppression of workers' reproduction?  We derive the following main results.

\begin{figure}
\centering
\begin{subfigure}{0.45\textwidth}
\centering
\caption{}
\includegraphics*[width=1\textwidth]{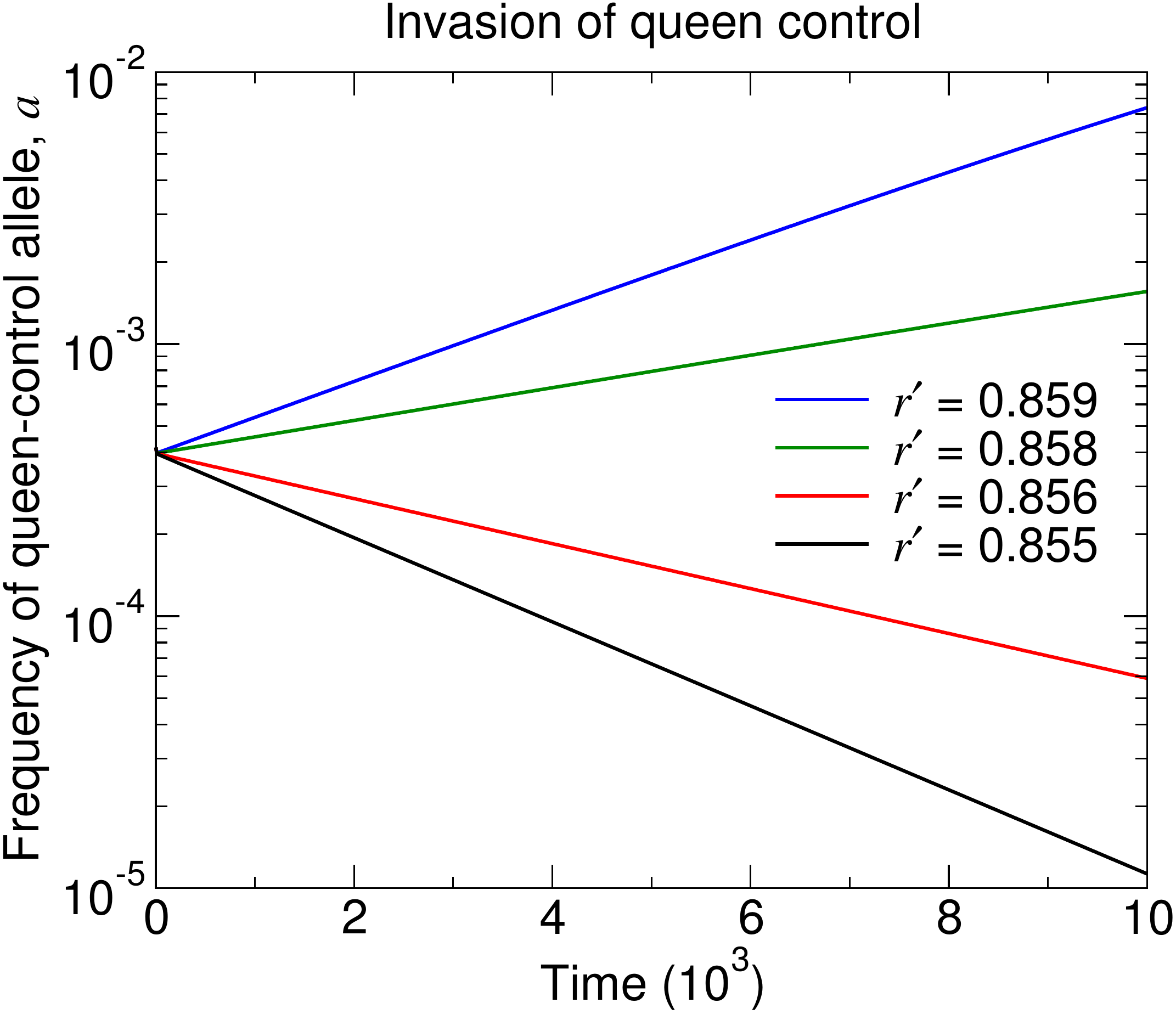}
\end{subfigure}
\quad
\begin{subfigure}{0.5\textwidth}
\centering
\caption{}
\includegraphics*[width=1\textwidth]{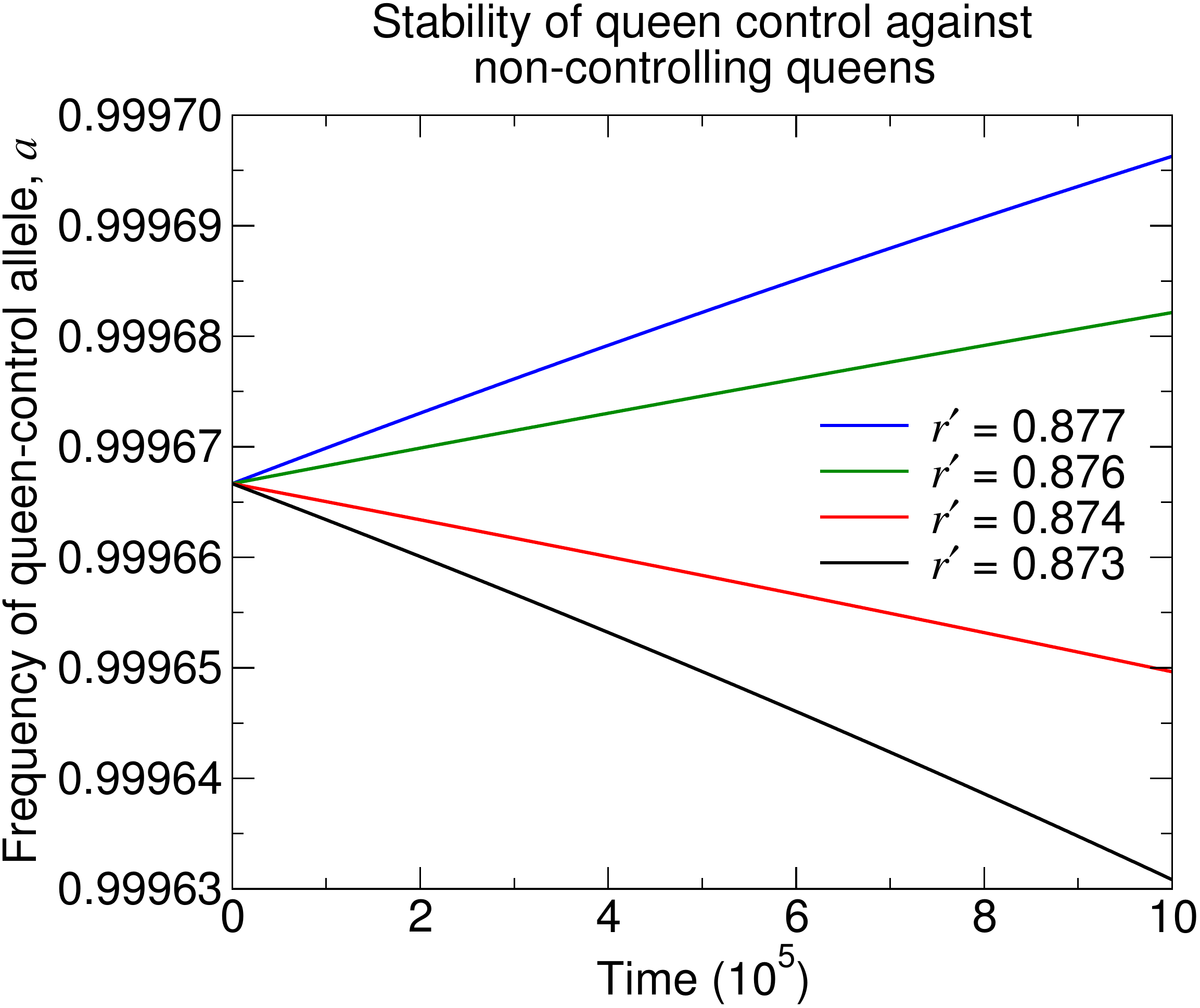}
\end{subfigure}
\caption{Numerical simulations demonstrate the condition for evolutionary invasion of queen control ((a), Condition \eqref{eqn:invasion}) and the condition for queen control to be evolutionarily stable, when fixed, against invasion by non-controlling queens ((b), Condition \eqref{eqn:stability}).  For these plots, we consider a dominant queen-control allele with singly mated queens ($n=1$), and we set $p=0.5$, $p'=1$, and $r=1$.  (The initial conditions are (a) $X_{AA,0}=1-10^{-3}$ and $X_{AA,1}=10^{-3}$ for each of the four curves, and (b) $X_{aa,1}=1-10^{-3}$ and $X_{aa,0}=10^{-3}$ for each of the four curves.)}
\label{fig:n1domsim}
\end{figure}

The $a$ allele, which causes the queen to suppress her workers' reproduction, invades a population of non-controlling queens if the following condition holds:
\begin{equation}
\frac{r'}{r} > \frac{2+2p}{2+p+p'}
\label{eqn:invasion}
\end{equation}
Condition \eqref{eqn:invasion} applies regardless of whether the queen-control allele, $a$, is dominant or recessive.  The evolutionary dynamics demonstrating Condition \eqref{eqn:invasion} for single mating and for a dominant queen-control allele are shown in Figure \ref{fig:n1domsim}(a).

Furthermore, the queen-control allele, $a$, when fixed in the population, is stable against invasion by the non-controlling $A$ allele if the following condition holds:
\begin{equation}
\frac{r'}{r} > \frac{2+p+p'}{2+2p'}
\label{eqn:stability}
\end{equation}
Condition \eqref{eqn:stability} also applies regardless of whether the queen-control allele, $a$, is dominant or recessive.  The evolutionary dynamics demonstrating Condition \eqref{eqn:stability} for single mating and for a dominant queen-control allele are shown in Figure \ref{fig:n1domsim}(b).

\begin{figure}
\begin{center}
\includegraphics[width=0.6\textwidth]{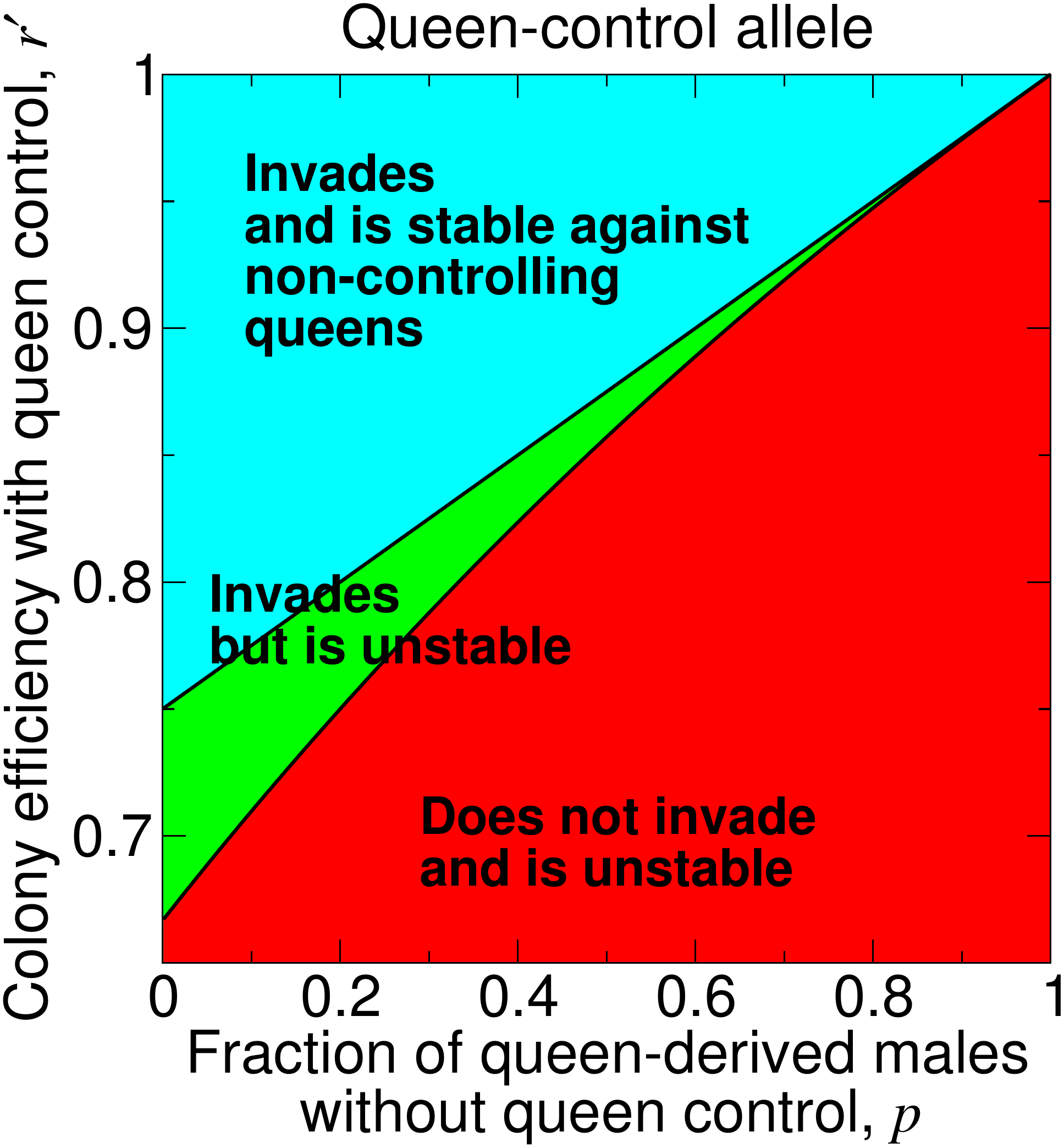}
\caption{A plot of $r'$ versus $p$ shows the three possibilities for the dynamical behavior of the queen-control allele around the two pure equilibria.  For this plot, we set $r=1$ and $p'=1$.}
\label{fig:regions}
\end{center}
\end{figure}

\begin{figure}
\centering
\begin{subfigure}{0.32\textwidth}
\centering
\caption{}
\includegraphics*[width=1\textwidth]{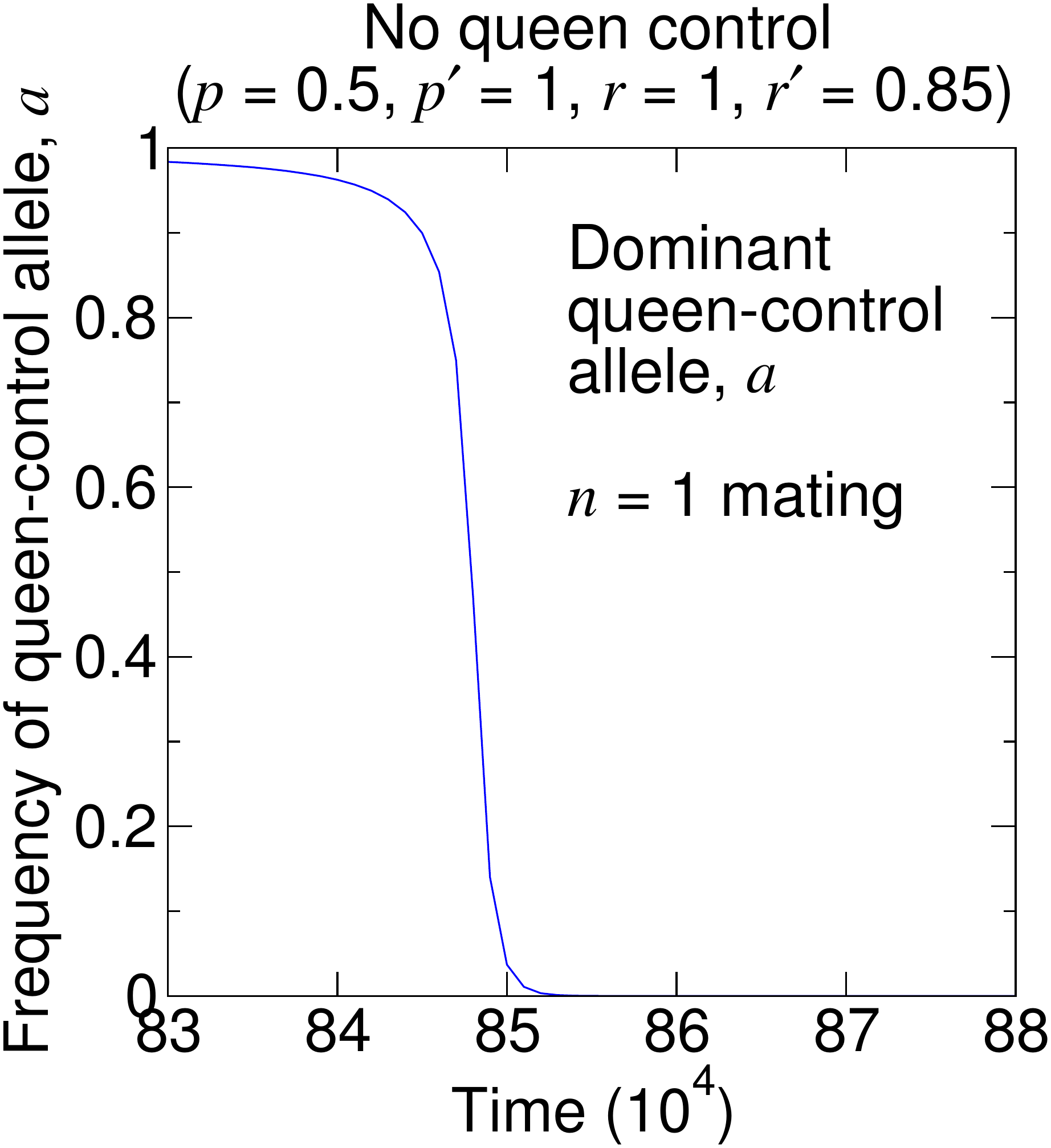}
\end{subfigure}
\begin{subfigure}{0.32\textwidth}
\centering
\caption{}
\includegraphics*[width=1\textwidth]{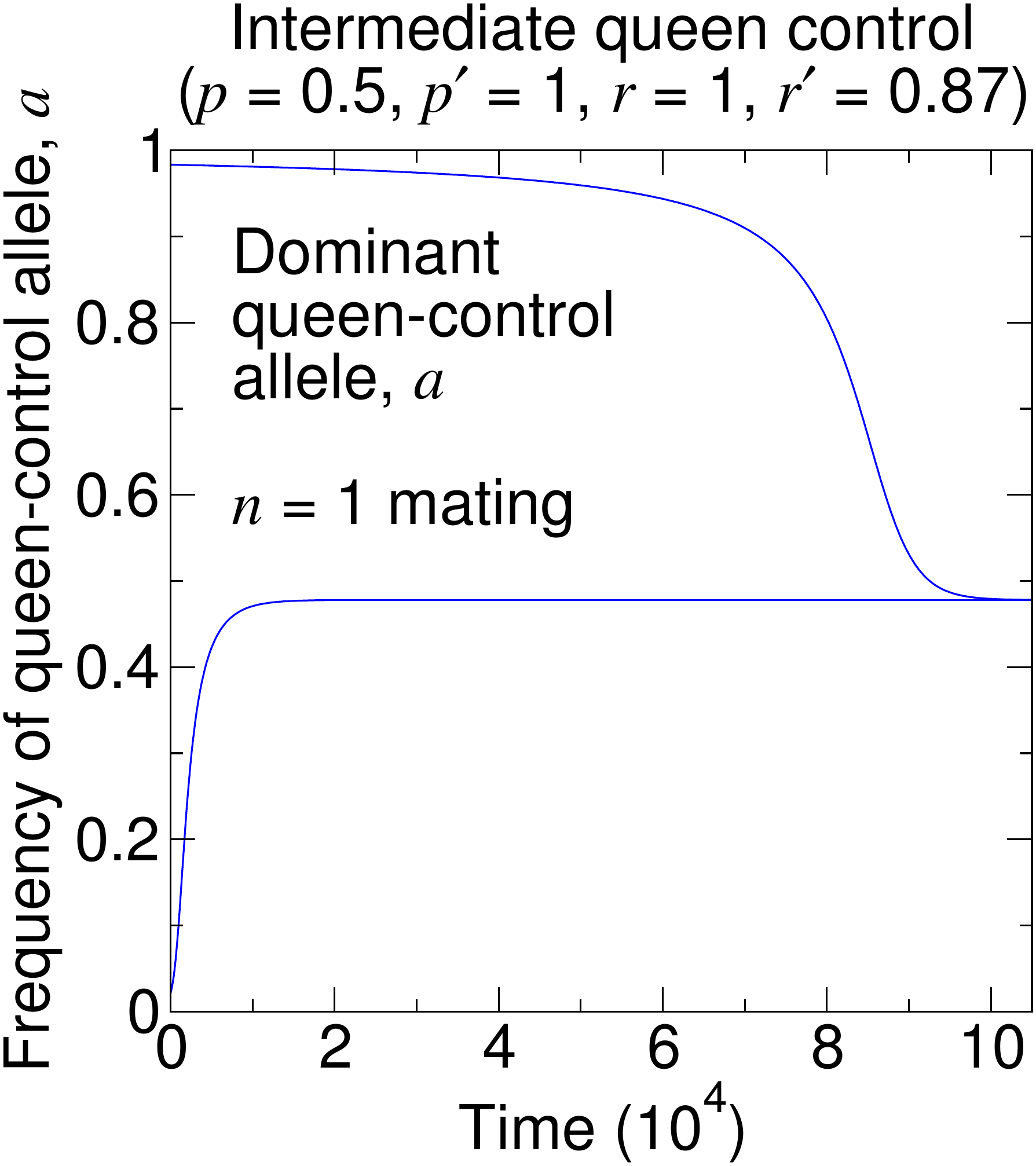}
\end{subfigure}
\begin{subfigure}{0.32\textwidth}
\centering
\caption{}
\includegraphics*[width=1\textwidth]{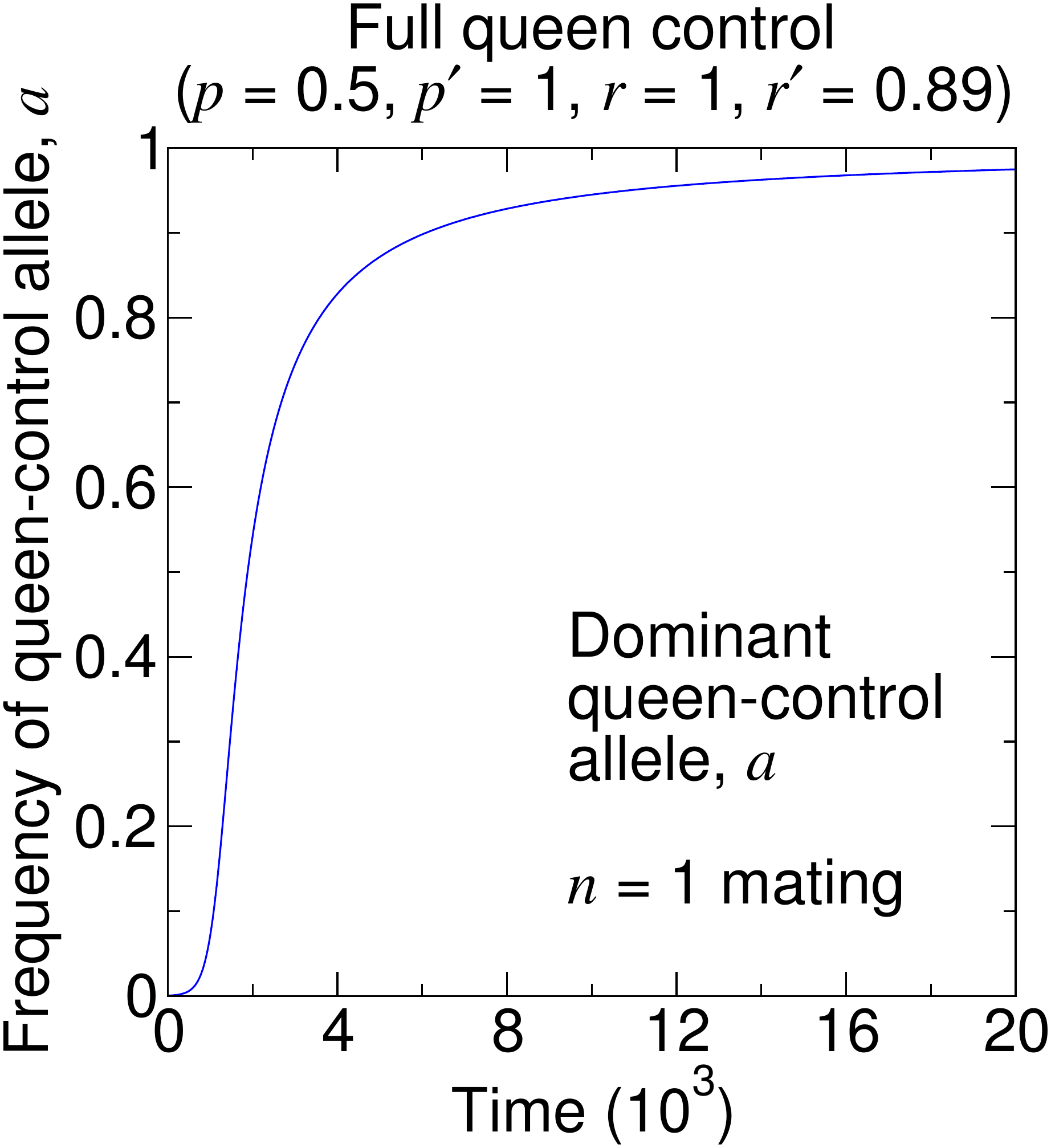}
\end{subfigure}
\caption{Simulations of the evolutionary dynamics show the dynamical behavior of the queen-control allele.  (a) corresponds to the lower, red region in Figure \ref{fig:regions}.  (b) corresponds to the middle, green region in Figure \ref{fig:regions}.  (c) corresponds to the upper, blue region in Figure \ref{fig:regions}.  (The initial conditions are (a) $X_{aa,1}=1-10^{-3}$ and $X_{aa,0}=10^{-3}$, (b, lower curve) $X_{AA,0}=1-0.05$ and $X_{AA,1}=0.05$, (b, upper curve) $X_{aa,1}=1-0.05$ and $X_{aa,0}=0.05$, and (c) $X_{AA,0}=1-10^{-3}$ and $X_{AA,1}=10^{-3}$.)}
\label{fig:regions_sim}
\end{figure}

Note that Condition \eqref{eqn:invasion} is always easier to satisfy than Condition \eqref{eqn:stability}.  Therefore, three scenarios regarding the two pure equilibria are possible:  The first possibility is that queen control is unable to invade a wild-type population and is unstable, when fixed, against invasion by non-control.  The second possibility is that queen control is able to invade a wild-type population but is unstable, when fixed, against invasion by non-control.  The third possibility is that queen control is able to invade a wild-type population and is stable, when fixed, against invasion by non-control.  In the case where queen control can invade a wild-type population but is unstable when fixed, Brouwer's fixed-point theorem guarantees the existence of at least one mixed equilibrium at which controlling and non-controlling queens coexist.  Regions of the parameter space are shown in Figure \ref{fig:regions}, and evolutionary dynamics illustrating the three scenarios are shown in Figure \ref{fig:regions_sim}.

Two salient points regarding the dynamics of the queen-control allele deserve emphasis.  First, the conditions for evolutionary invasion and stability of queen control do not depend on the queen's mating number $n$.  To develop intuition, consider the introduction of an initially rare dominant allele for queen control.  When the allele is rare, for $n$ matings, $AA,1$ colonies are more abundant than $Aa,0$ colonies by a factor of $n$.  A fraction $(n-1)/n$ of offspring of $AA,1$ colonies arise from selecting sperm from wild-type males and are 100\% wild-type, as though they had originated from $AA,0$ colonies.  However, the remaining fraction $1/n$ of offspring of $AA,1$ colonies are produced in the same relative mutant/wild-type proportions as if they had originated from $AA,n$ colonies.  Notice that the factor of $n$ from the matings cancels with the probability of $1/n$ of selecting sperm from the mutant male.  Therefore, we have a simple interpretation:  For considering invasion of queen control, and at the leading-order frequency of the mutant allele, the system effectively consists of $AA,n$ colonies and $Aa,0$ colonies at relative amounts that do not depend on $n$.  But $AA,n$ colonies produce mutant and wild-type offspring in relative proportions that do not depend on $n$, and $Aa,0$ colonies produce mutant and wild-type offspring in relative proportions that do not depend on $n$.  Thus, $n$ does not enter into Condition \eqref{eqn:invasion}.

Second, queen control can evolve even if it results in efficiency losses.  To see why, notice that, if the queen-control allele is dominant, then type $Aa,0$ colonies have the mutant phenotype, and if the queen-control allele is recessive, then type $aa,0$ colonies have the mutant phenotype.  In the dominant case, workers in type $Aa,0$ colonies produce $3$ type $A$ males for every type $a$ male, but the queen produces type $A$ and type $a$ males in equal proportion.  In the recessive case, workers in type $aa,0$ colonies produce type $A$ and type $a$ males in equal proportion, but the queen produces only type $a$ males.  In both cases, if the queen takes over production of males (i.e., if $p'>p$), then the frequency of the mutant allele in the next generation increases.

Thus, the allele for queen control can act as a selfish genetic element, enabling queen-induced worker sterility to develop in a population even if it diminishes colony reproductive efficiency.  Queens are easily selected to increase their production of male offspring and suppress workers' production of male offspring. In this case, workers might also be selected to evade manipulation by queens, setting up an evolutionary arms race.  When does queen control evolve and persist in the population?

Consider the following scenario.  Initially, there is a homogeneous population of colonies.  All queens are homozygous for allele $A$ at locus $\mathcal{A}$, and all workers are homozygous for allele $B$ at locus $\mathcal{B}$.  In each colony, the fraction of queen-derived males within the colony is $p$, and the overall reproductive efficiency of the colony is $r$.  Suppose that a mutation, $a$, acts in a queen at locus \emph{A}, causing her to completely suppress her workers' production of drones.  In colonies headed by controlling queens, all males originate from the controlling queen ($p'=1$), and the overall reproductive efficiency of the colony is $r'$.  According to Equations \eqref{eqn:invasion} and \eqref{eqn:stability}, if $r'/r$ is sufficiently large ($>[3+p]/4$), then the controlling queens will increase in frequency and fix in the population.  Once the queen-control allele has fixed, each colony's male eggs originate only from the queen ($p'=1$), and each colony has overall reproductive efficiency $r'$.

Next, consider a subsequent mutation, $b$, that acts in workers at locus $\mathcal{B}$.  The $b$ allele changes a worker's phenotype, causing the mutant worker to become reproductive again.  The $b$ allele for worker reproduction can be either dominant, so that type $Bb$ and type $bb$ workers are reproductive, or recessive, so that only type $bb$ workers are reproductive \citep{Olejarz_2015}.  If a colony contains only workers with the reproductive phenotype, then the fraction of queen-derived males within the colony is $p$, and the overall reproductive efficiency of the colony is $r$.  Thus, the $b$ allele for worker reproduction essentially undoes the effects of the $a$ allele for queen control.

What are the requirements for queen control to be evolutionarily stable against a mutation in workers that restores their reproduction?  To answer this question for a dominant $b$ allele, we turn to Equation (53) in \cite{Olejarz_2015}, which is the condition, for any number of matings, $n$, for stability of a recessive mutation in workers that results in worker sterility:  Setting $r_1=r'$ in Equation (53) in \cite{Olejarz_2015}, this condition becomes
\begin{equation}
\left[ \frac{r'}{r_\frac{n-1}{n}} - \frac{n\left(1-p_\frac{n-1}{n}\right)}{2} \right] \left[ 2\left(\frac{r'}{r_\frac{1}{2}}\right) - 1 \right] > 1
\label{eqn:resistance_rec_ster_sta}
\end{equation}
In Condition \eqref{eqn:resistance_rec_ster_sta}, $r_{1/2}$ is the colony reproductive efficiency when a fraction $1/2$ of workers are reproductive, $r_{(n-1)/n}$ is the colony reproductive efficiency when a fraction $1/n$ of workers are reproductive, and $p_{(n-1)/n}$ is the fraction of queen-derived males when a fraction $1/n$ of workers are reproductive.  If Condition \eqref{eqn:resistance_rec_ster_sta} is satisfied, then a subsequent dominant mutation, $b$, that acts in workers to restore their reproduction \emph{cannot} invade a queen-controlled population.
 
To further determine if the dominant $b$ allele cannot fix, we must also consider the equation directly after Equation (34) in \cite{Olejarz_2015}, which is the condition, for any number of matings, $n$, for invasion of a recessive mutation in workers that results in worker sterility.  Setting $p_0=p$ and $r_0=r$ in the equation directly after Equation (34) in \cite{Olejarz_2015}, we obtain
\begin{equation}
\frac{r_\frac{1}{2n}}{r} > \frac{2(2n-1)(2+n+np)}{2n^2\left(2+p+p_\frac{1}{2n}\right)+n\left(3+3p-2p_\frac{1}{2n}\right)-2(1+p)}
\label{eqn:resistance_rec_ster_inv}
\end{equation}
In Condition \eqref{eqn:resistance_rec_ster_inv}, $r_{1/(2n)}$ is the colony reproductive efficiency when a fraction $(2n-1)/(2n)$ of workers are reproductive, and $p_{1/(2n)}$ is the fraction of queen-derived males when a fraction $(2n-1)/(2n)$ of workers are reproductive.  If Condition \eqref{eqn:resistance_rec_ster_inv} is satisfied, then a subsequent dominant mutation, $b$, that acts in workers to restore their reproduction \emph{cannot} fix in the population.

Notice that Condition \eqref{eqn:resistance_rec_ster_sta} depends on the parameters $r_{1/2}$, $r_{(n-1)/n}$, and $p_{(n-1)/n}$, which are related to the effects of the $b$ allele for worker reproduction.  Also, notice that Condition \eqref{eqn:resistance_rec_ster_inv} depends on the parameters $r_{1/(2n)}$ and $p_{1/(2n)}$, which are related to the effects of the $b$ allele for worker reproduction.  The properties of the particular dominant $b$ allele for worker reproduction that is under consideration are therefore essential for determining if the effects of the $a$ allele for queen control can be undone by worker resistance.

To gain insight, regarding the parameters $r_{1/2}$, $r_{(n-1)/n}$, $p_{(n-1)/n}$, $r_{1/(2n)}$, and $p_{1/(2n)}$ in Conditions \eqref{eqn:resistance_rec_ster_sta} and \eqref{eqn:resistance_rec_ster_inv}, we can consider the following simple case:
\begin{equation}
\begin{aligned}
p_z &= p+(1-p)z \\
r_z &= r+(r'-r)z \\
\end{aligned}
\label{eqn:parameters_1}
\end{equation}
For the parameter choices given by Equations \eqref{eqn:parameters_1}, Condition \eqref{eqn:resistance_rec_ster_sta} becomes
\begin{equation}
\frac{r'}{r} > \frac{4(1-p)+n(3+p)+\sqrt{4(3-p)^2+n^2(3+p)^2+4n(7+p^2)}}{10+2n+6(n-1)p}
\label{eqn:resistance_rec_ster_sta_particular}
\end{equation}
Also for the parameter choices given by Equations \eqref{eqn:parameters_1}, Condition \eqref{eqn:resistance_rec_ster_inv} becomes
\begin{equation}
\frac{r'}{r} > \frac{3+4n+p}{3+p+2n(1+p)}
\label{eqn:resistance_rec_ster_inv_particular}
\end{equation}

To determine if queen control is evolutionarily stable against a recessive $b$ mutation in workers that restores their reproduction, we turn to the equation directly after Equation (49) in \cite{Olejarz_2015}, which is the condition, for any number of matings, $n$, for stability of a dominant mutation in workers that results in worker sterility:  Setting $r_1=r'$ in the equation directly after Equation (49) in \cite{Olejarz_2015}, this condition becomes
\begin{equation}
\frac{r'}{r_\frac{2n-1}{2n}} > \frac{2+3n-np_\frac{2n-1}{2n}}{2(n+1)}
\label{eqn:resistance_dom_ster_sta}
\end{equation}
In Condition \eqref{eqn:resistance_dom_ster_sta}, $r_{(2n-1)/(2n)}$ is the colony reproductive efficiency when a fraction $1/(2n)$ of workers are reproductive, and $p_{(2n-1)/(2n)}$ is the fraction of queen-derived males when a fraction $1/(2n)$ of workers are reproductive.  If Condition \eqref{eqn:resistance_dom_ster_sta} is satisfied, then a subsequent recessive mutation, $b$, that acts in workers to restore their reproduction \emph{cannot} invade a queen-controlled population.
 
To further determine if the recessive $b$ allele cannot fix, we must also consider Equation (20) in \cite{Olejarz_2015}, which is the condition, for any number of matings, $n$, for invasion of a dominant mutation in workers that results in worker sterility.  Setting $r_0=r$ in Equation (20) in \cite{Olejarz_2015}, we obtain
\begin{equation}
\frac{r_\frac{1}{2}}{r} \left[ 1 + p_\frac{1}{2}\left(\frac{r_\frac{1}{n}}{r}\right) \right] > 2
\label{eqn:resistance_dom_ster_inv}
\end{equation}
In Condition \eqref{eqn:resistance_dom_ster_inv}, $r_{1/n}$ is the colony reproductive efficiency when a fraction $(n-1)/n$ of workers are reproductive, $r_{1/2}$ is the colony reproductive efficiency when a fraction $1/2$ of workers are reproductive, and $p_{1/2}$ is the fraction of queen-derived males when a fraction $1/2$ of workers are reproductive.  If Condition \eqref{eqn:resistance_dom_ster_inv} is satisfied, then a subsequent recessive mutation, $b$, that acts in workers to restore their reproduction \emph{cannot} fix in the population.

Notice that Condition \eqref{eqn:resistance_dom_ster_sta} depends on the parameters $r_{(2n-1)/(2n)}$ and $p_{(2n-1)/(2n)}$, which are related to the effects of the $b$ allele for worker reproduction.  Also, notice that Condition \eqref{eqn:resistance_dom_ster_inv} depends on the parameters $r_{1/n}$, $r_{1/2}$, and $p_{1/2}$, which are related to the effects of the $b$ allele for worker reproduction.  The properties of the particular recessive $b$ allele for worker reproduction that is under consideration are therefore essential for determining if the effects of the $a$ allele for queen control can be undone by worker resistance.

To gain insight, regarding the parameters $r_{(2n-1)/(2n)}$, $p_{(2n-1)/(2n)}$, $r_{1/n}$, $r_{1/2}$, and $p_{1/2}$ in Conditions \eqref{eqn:resistance_dom_ster_sta} and \eqref{eqn:resistance_dom_ster_inv}, we can again consider the simple case given by Equations \eqref{eqn:parameters_1}.  For the parameter choices given by Equations \eqref{eqn:parameters_1}, Condition \eqref{eqn:resistance_dom_ster_sta} becomes
\begin{equation}
\frac{r'}{r} > \frac{5+4n-p}{5-p+2n(1+p)}
\label{eqn:resistance_dom_ster_sta_particular}
\end{equation}
Also for the parameter choices given by Equations \eqref{eqn:parameters_1}, Condition \eqref{eqn:resistance_dom_ster_inv} becomes
\begin{equation}
\frac{r'}{r} > \frac{\sqrt{4n(5-p)(1+p)+4(1+p)^2+n^2(3+p)^2}-n(3+p)}{2(1+p)}
\label{eqn:resistance_dom_ster_inv_particular}
\end{equation}

\begin{figure}
\centering
\begin{subfigure}{0.35\textwidth}
\centering
\caption{}
\includegraphics*[width=1\textwidth]{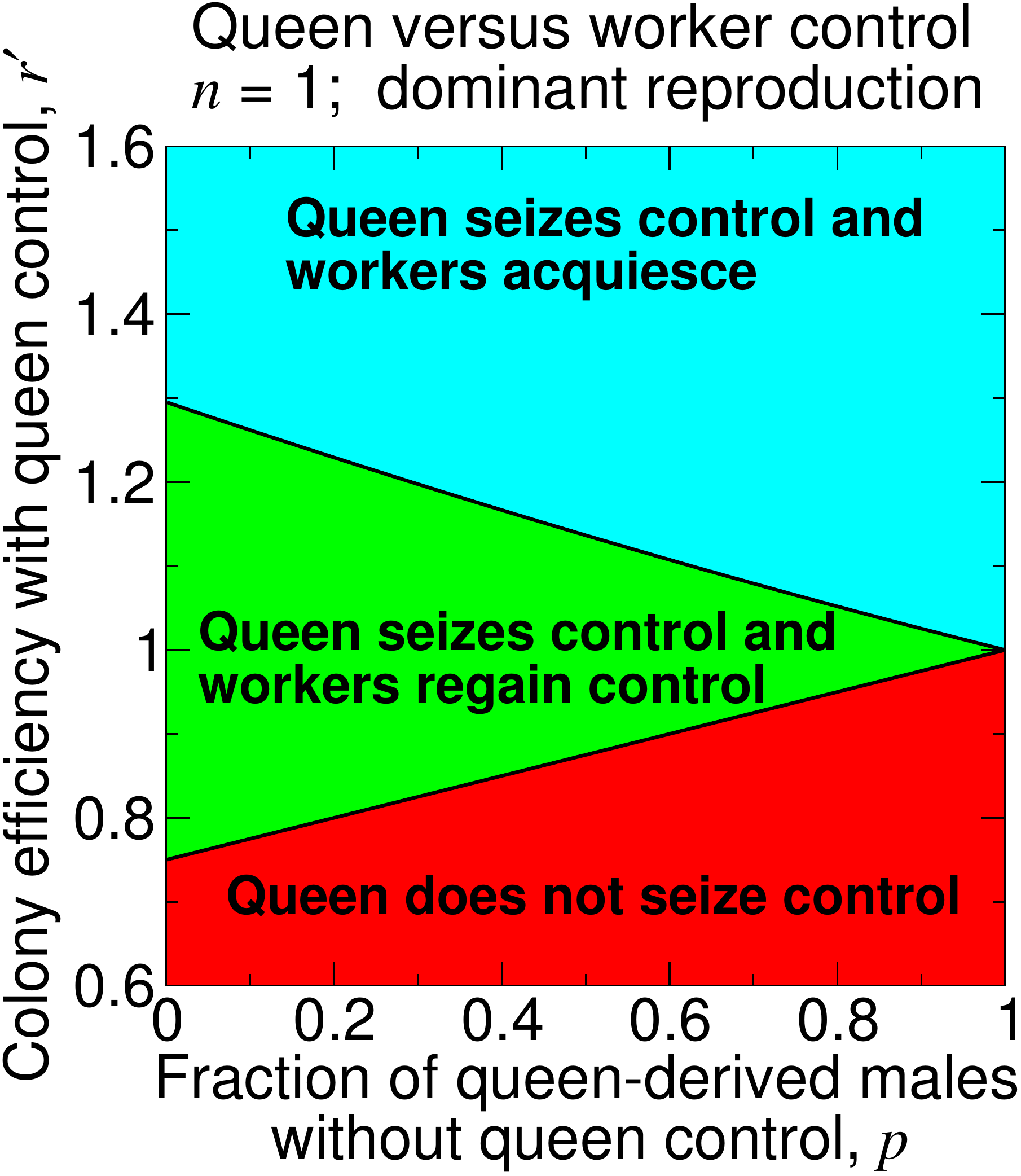}
\end{subfigure}
\quad
\begin{subfigure}{0.35\textwidth}
\centering
\caption{}
\includegraphics*[width=1\textwidth]{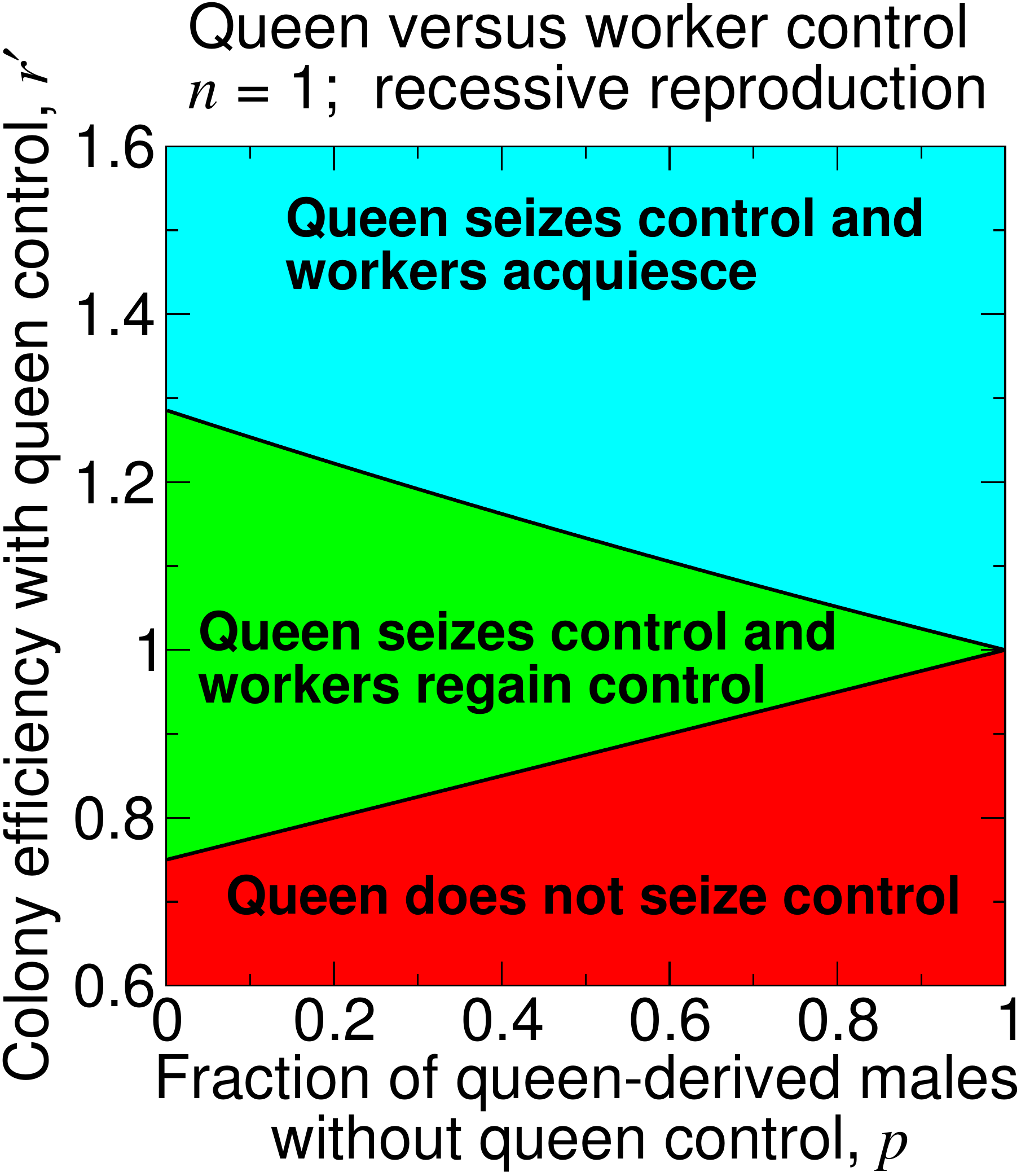}
\end{subfigure}
\par\medskip
\begin{subfigure}{0.35\textwidth}
\centering
\caption{}
\includegraphics*[width=1\textwidth]{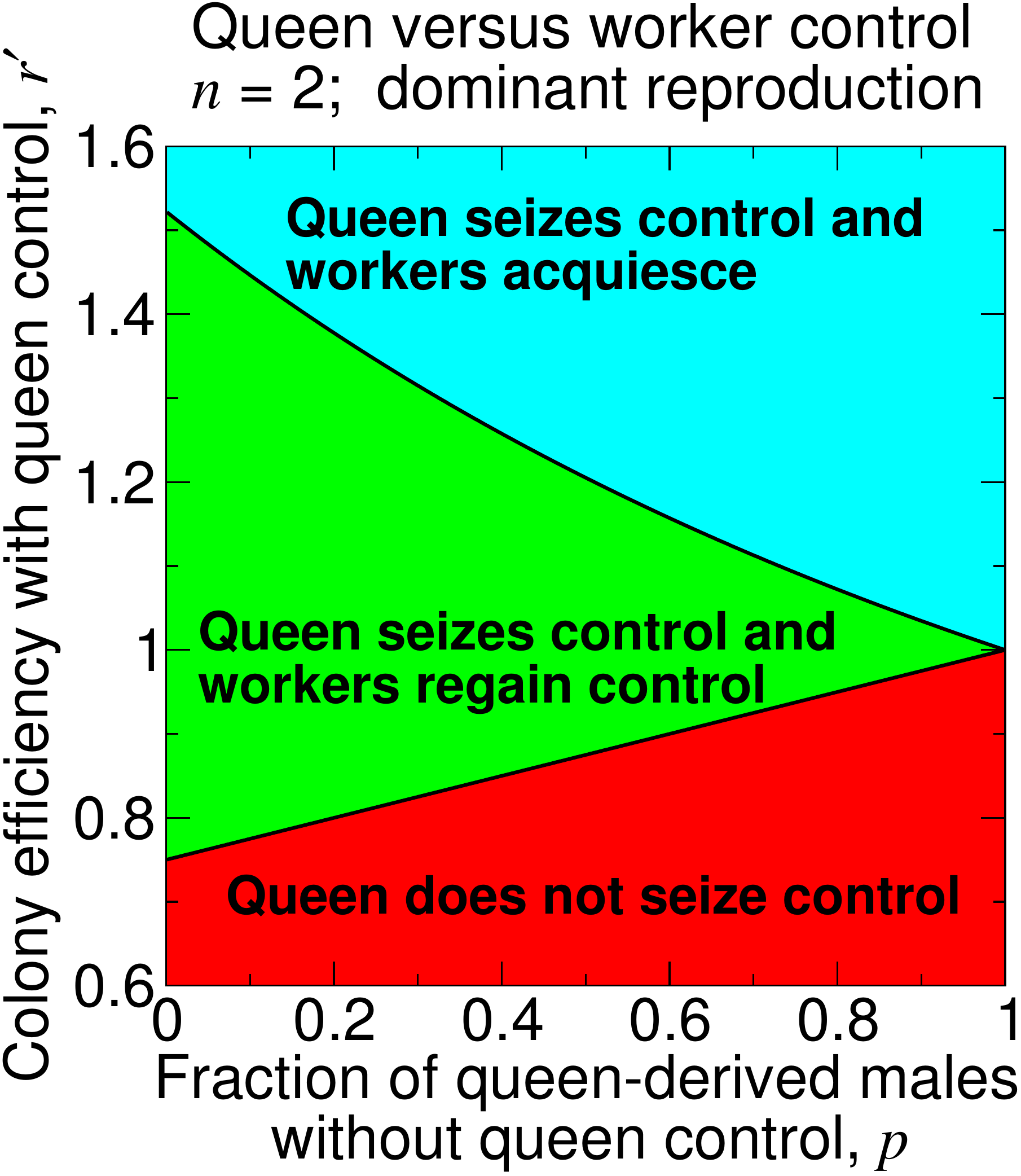}
\end{subfigure}
\quad
\begin{subfigure}{0.35\textwidth}
\centering
\caption{}
\includegraphics*[width=1\textwidth]{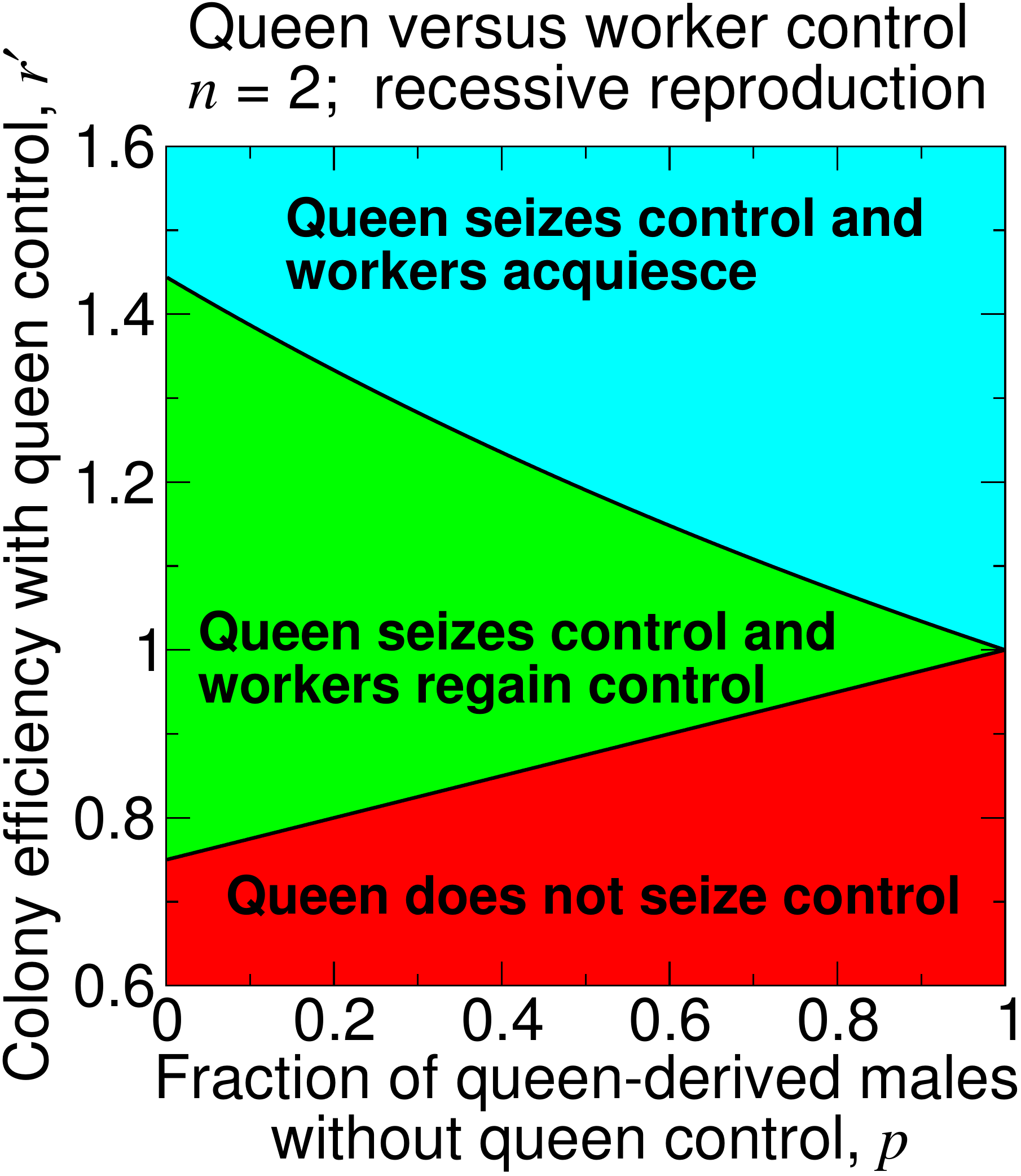}
\end{subfigure}
\caption{A mutation for queen control may or may not be in conflict with workers' reproductive capabilities.  We set $r=1$ without loss of generality, and we assume that the queen-control allele eliminates workers' reproduction.  If the efficiency loss from queen control is too severe (corresponding to values of $r'$ in the red region), then queen control does not evolve (or it invades without fixing, and a subsequent mutation acting in workers causes them to become fully reproductive again).  If the efficiency loss or gain from queen control is moderate (corresponding to values of $r'$ in the green region), then queen control evolves, but a subsequent mutation acting in workers causes them to become fully reproductive again.  If the efficiency gain from queen control is sufficiently large (corresponding to values of $r'$ in the blue region), then queen control evolves, and workers subsequently acquiesce by remaining non-reproductive.  The lower boundary is given by Equation \eqref{eqn:stability}, and the upper boundary is given by (a) Equation \eqref{eqn:resistance_rec_ster_sta_particular} for $n=1$, (b) Equation \eqref{eqn:resistance_dom_ster_sta_particular} for $n=1$, (c) Equation \eqref{eqn:resistance_rec_ster_sta_particular} for $n=2$, and (d) Equation \eqref{eqn:resistance_dom_ster_sta_particular} for $n=2$.  For this plot, we use Equations \eqref{eqn:parameters_1}, and we set $p'=1$ and $r=1$.}
\label{fig:rep}
\end{figure}

Figure \ref{fig:rep} shows the evolutionary outcome of queen control for parameters $p$ and $r'$.  We set $r=1$ without loss of generality.  In each panel, the boundary between the lower, red region and the middle, green region is given by Condition \eqref{eqn:stability}.  The boundary between the middle, green region and the upper, blue region is given by Condition \eqref{eqn:resistance_rec_ster_sta_particular} for $n=1$ (Figure \ref{fig:rep}(a)), Condition \eqref{eqn:resistance_dom_ster_sta_particular} for $n=1$ (Figure \ref{fig:rep}(b)), Condition \eqref{eqn:resistance_rec_ster_sta_particular} for $n=2$ (Figure \ref{fig:rep}(c)), and Condition \eqref{eqn:resistance_dom_ster_sta_particular} for $n=2$ (Figure \ref{fig:rep}(d)).  For values $(p,r')$ in the lower, red region, the $a$ mutation for queen control is unable to spread to fixation.  For values $(p,r')$ in the middle, green region, the $a$ mutation for queen control invades and is evolutionarily stable to non-control, but the subsequent $b$ mutation for worker reproduction also invades and is evolutionarily stable, undoing the effects of queen control.  For values $(p,r')$ in the upper, blue region, the $a$ mutation for queen control invades and is evolutionarily stable to non-control, and the subsequent $b$ mutation for worker reproduction is unable to invade, rendering queen control evolutionarily stable against counteraction by workers.

\begin{figure}
\centering
\begin{subfigure}{0.45\textwidth}
\centering
\caption{}
\includegraphics*[width=1\textwidth]{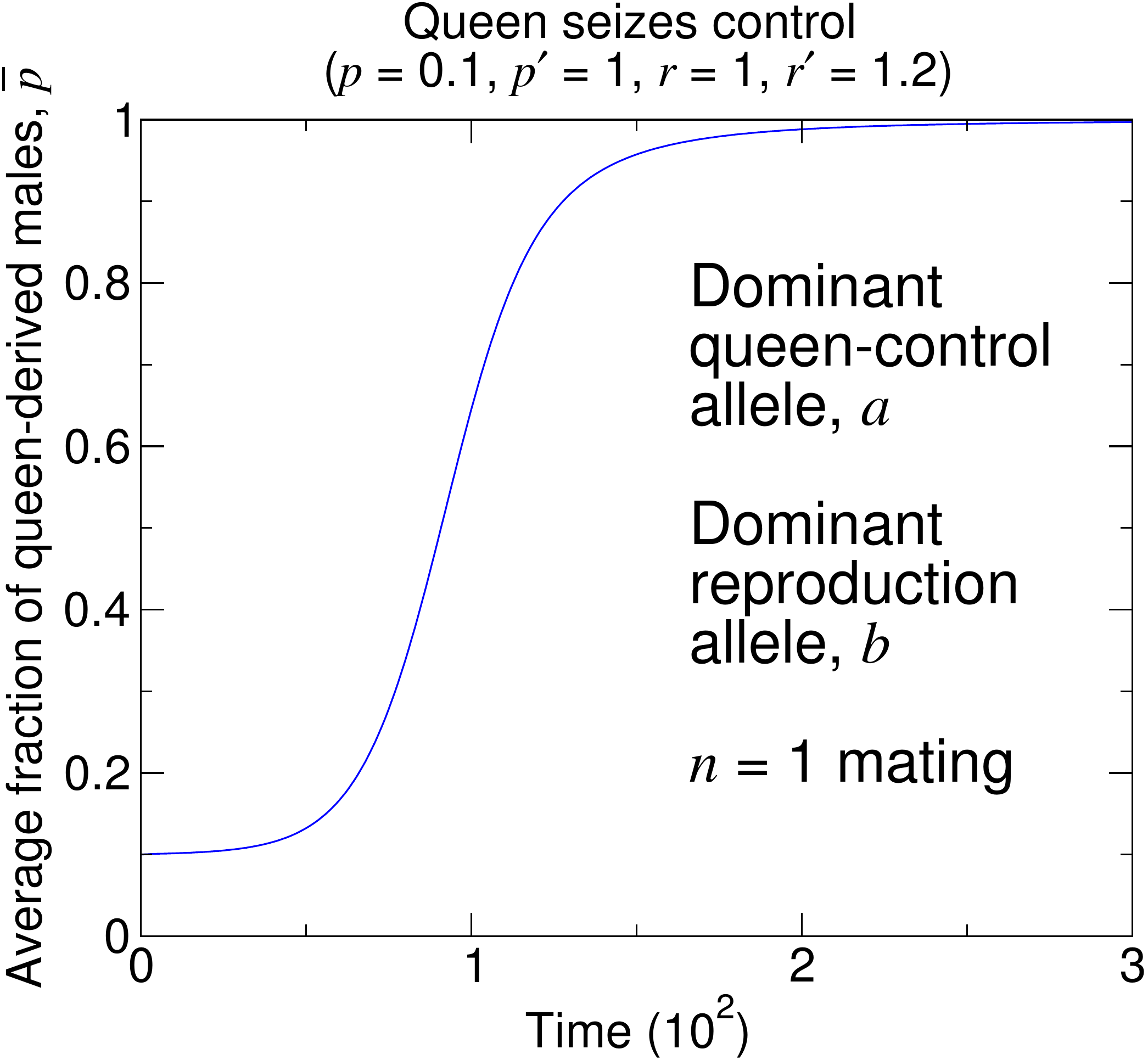}
\end{subfigure}
\quad
\begin{subfigure}{0.45\textwidth}
\centering
\caption{}
\includegraphics*[width=1\textwidth]{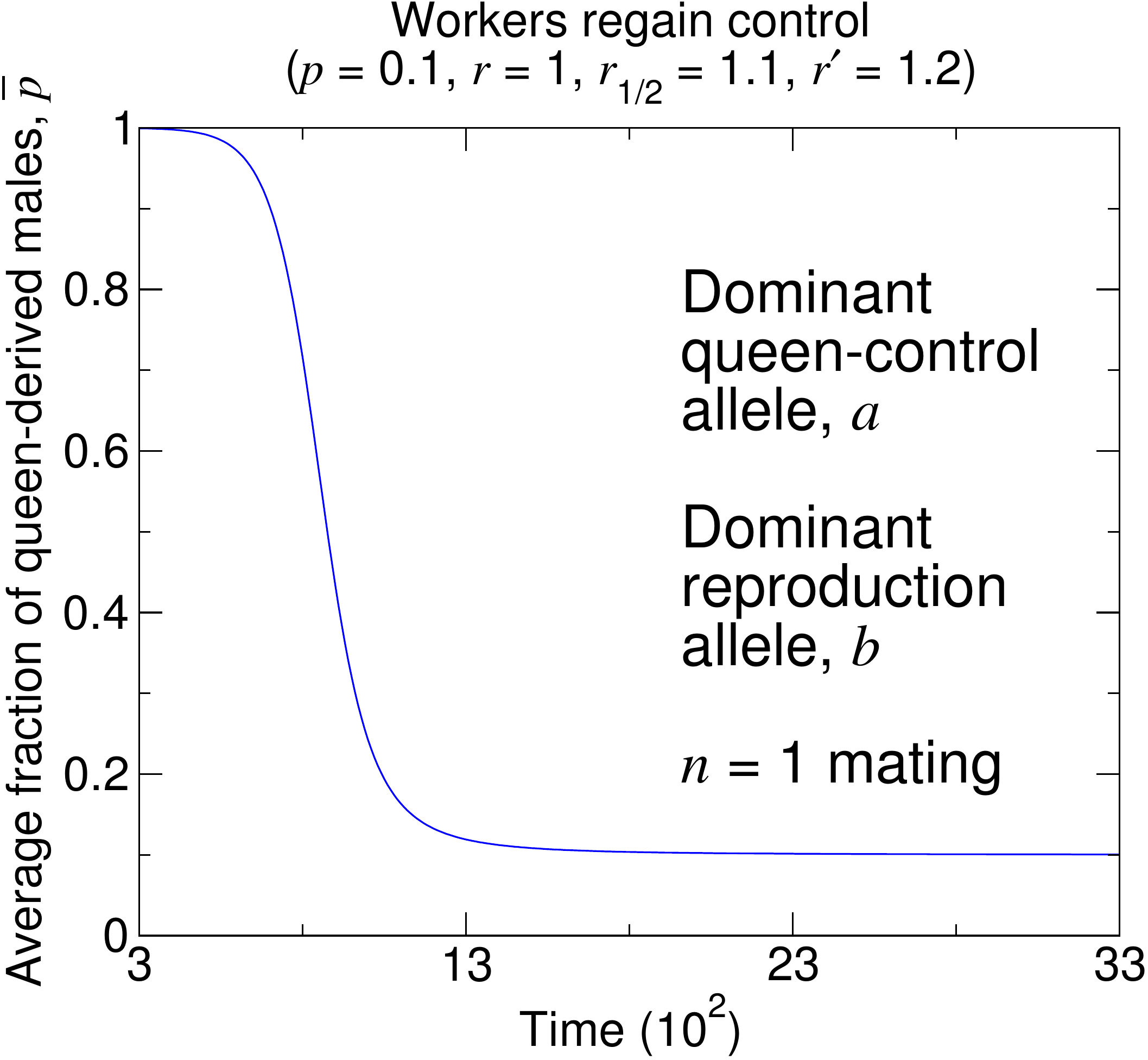}
\end{subfigure}
\par\medskip
\begin{subfigure}{0.45\textwidth}
\centering
\caption{}
\includegraphics*[width=1\textwidth]{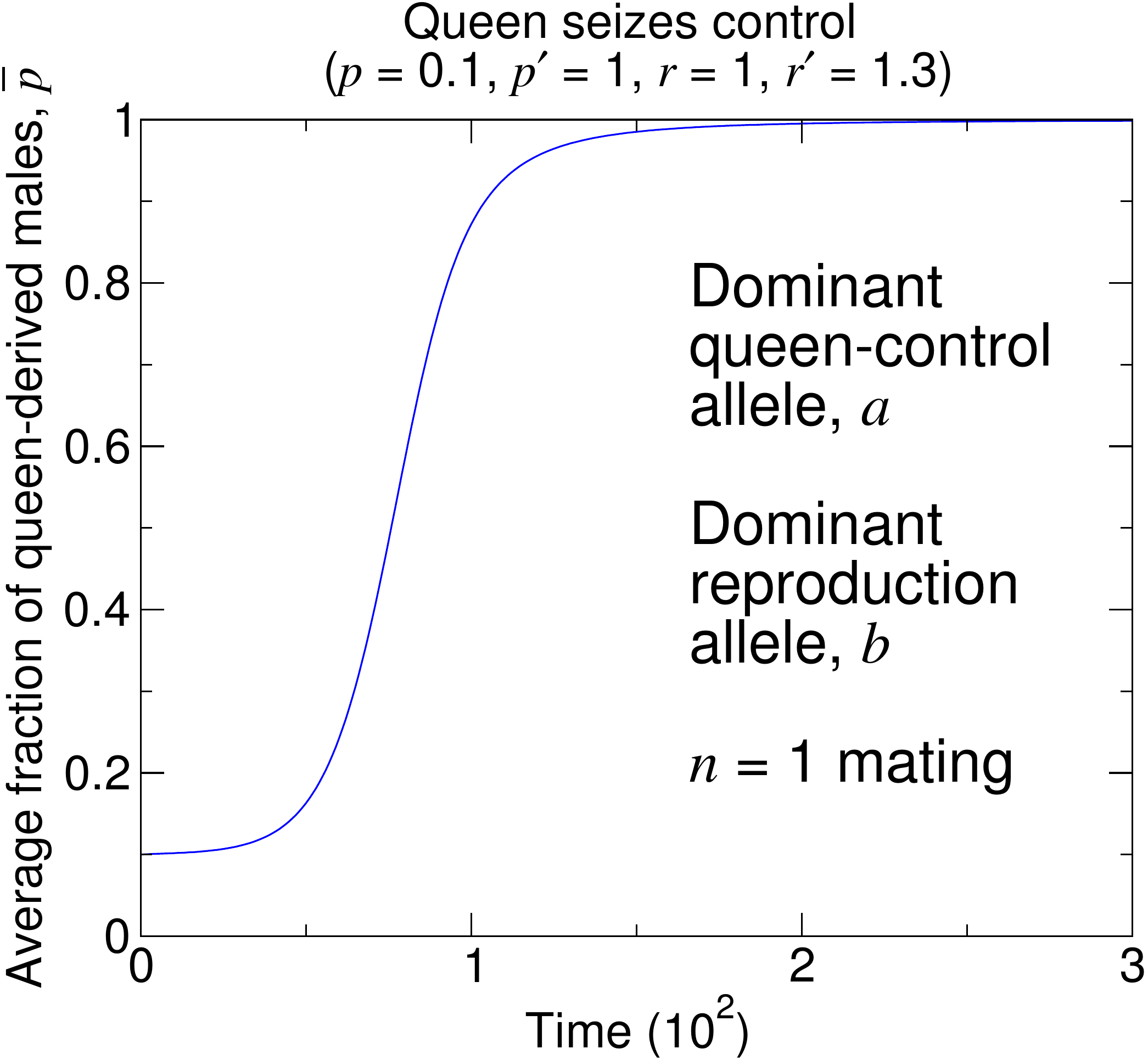}
\end{subfigure}
\quad
\begin{subfigure}{0.45\textwidth}
\centering
\caption{}
\includegraphics*[width=1\textwidth]{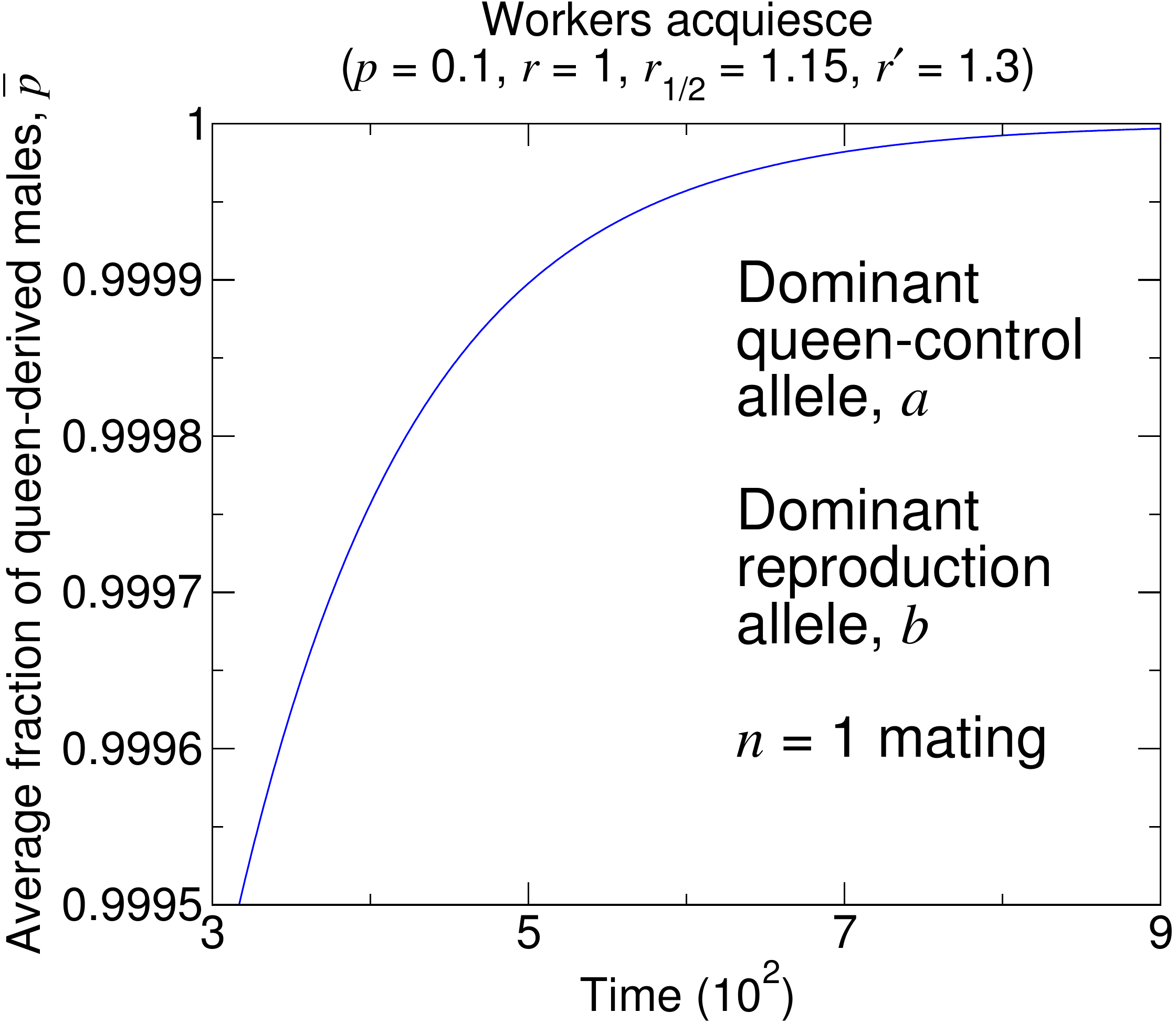}
\end{subfigure}
\caption{Simulations reveal the behaviors shown in Figure \ref{fig:rep}.  It's possible that a mutation causing queen control evolves (a), and worker reproduction is subsequently restored (b).  But if the efficiency gain due to queen control is large enough, then queen control evolves (c), and workers are unable to regain reproductive ability (d).  (In (b) and (d), $r_{1/2}$ denotes the colony efficiency when $1/2$ of workers in the colony have the phenotype for worker reproduction.  We follow the assumption for $r_z$ in Equations \eqref{eqn:parameters_1} for determining the values of $r_{1/2}=1.1$ (b) and $r_{1/2}=1.15$ (d) for these simulations.  The initial conditions are (a) $X_{AA,0}=1-10^{-3}$ and $X_{AA,1}=10^{-3}$, (b) $X_{BB,0}=1-10^{-3}$ and $X_{BB,1}=10^{-3}$, (c) $X_{AA,0}=1-10^{-3}$ and $X_{AA,1}=10^{-3}$, and (d) $X_{BB,0}=1-10^{-3}$ and $X_{BB,1}=10^{-3}$.  For (b) and (d), we introduce the $b$ allele for worker reproduction at time $t=300$.)}
\label{fig:rep_sim}
\end{figure}

Corresponding simulations of the evolutionary dynamics are shown in Figure \ref{fig:rep_sim}.  In Figure \ref{fig:rep_sim}, the quantity $\overline{p}$ that is plotted on the vertical axis is the average fraction of queen-derived males in the population.  Since Figure \ref{fig:rep_sim} is for single mating ($n=1$) and a dominant queen-control allele, we have $\overline{p}=p(X_{AA,0}+X_{AA,1})+p'(X_{Aa,0}+X_{Aa,1}+X_{aa,0}+X_{aa,1})$, where $X_{AA,0}$, $X_{AA,1}$, $X_{Aa,0}$, $X_{Aa,1}$, $X_{aa,0}$, and $X_{aa,1}$ are the frequencies of the six types of colonies in the population.

There is a subtlety, however.  Figure \ref{fig:rep} assumes that queen control can be easily undone by a single mutation in workers.  This assumption is not necessarily true.  A single mutation in a worker may not be sufficient to reverse the primer or releaser effects of a queen's complex pheromonal bouquet.  The queen or dominant individual can also perform oophagy of worker-laid eggs or physical aggression, and it is unclear if a single mutation in a worker can enable her to overcome such behavioral dominance activities.

Thus, there is another important aspect to the question of evolutionary stability of queen control.  If there is a high genetic barrier against workers' resistance to partial queen control, then can \emph{partial} queen control incentivize workers to become completely sterile?

Consider, again, that there is initially a homogeneous population of colonies.  All queens are homozygous for allele $A$ at locus $\mathcal{A}$, and all workers are homozygous for allele $C$ at locus $\mathcal{C}$.  Each colony's fraction of queen-derived males is $p$, and each colony's overall reproductive efficiency is $r$.  Suppose that a mutation, $a$, acts in a queen at locus \emph{A}, causing her to partially suppress her workers' production of drones.  In colonies headed by partially controlling queens, a fraction $p'$ of males originate from the partially controlling queen, with $p<p'<1$, and the overall reproductive efficiency of the colony is $r'$.  According to Equations \eqref{eqn:invasion} and \eqref{eqn:stability}, if $r'/r$ is sufficiently large, then the partially controlling queens will increase in frequency and fix in the population.  Once the allele for partial queen control has fixed, a fraction $p'$ of each colony's male eggs originate from the queen, and each colony has overall reproductive efficiency $r'$.

Next, consider a subsequent mutation, $c$, that acts in workers at locus $\mathcal{C}$.  The $c$ allele changes a worker's phenotype, causing the mutant worker to become completely sterile.  The $c$ allele for worker sterility can be either recessive, so that only type $cc$ workers are sterile, or dominant, so that type $Cc$ and type $cc$ workers are sterile \citep{Olejarz_2015}.  If a colony contains only workers with the phenotype for sterility, then the fraction of queen-derived males within the colony is $1$, and the overall reproductive efficiency of the colony is $r^*$.

What are the requirements for partial queen control to enable the evolutionary success of a mutation in workers that renders them sterile?  To answer this question for a recessive $c$ allele, we turn to the equation directly after Equation (34) in \cite{Olejarz_2015}, which is the condition, for any number of matings, $n$, for invasion of a recessive mutation in workers that causes worker sterility:  Setting $p_0=p'$ and $r_0=r'$ in the equation directly after Equation (34) in \cite{Olejarz_2015}, this condition becomes
\begin{equation}
\frac{r_\frac{1}{2n}}{r'} > \frac{2(2n-1)(2+n+np')}{2n^2\left(2+p'+p_\frac{1}{2n}\right)+n\left(3+3p'-2p_\frac{1}{2n}\right)-2(1+p')}
\label{eqn:sterility_rec_ster_inv}
\end{equation}
In Condition \eqref{eqn:sterility_rec_ster_inv}, $r_{1/(2n)}$ is the colony reproductive efficiency when a fraction $1/(2n)$ of workers are sterile, and $p_{1/(2n)}$ is the fraction of queen-derived males when a fraction $1/(2n)$ of workers are sterile.  If Condition \eqref{eqn:sterility_rec_ster_inv} is satisfied, then a subsequent recessive mutation, $c$, that acts in workers to render them sterile invades a partially queen-controlled population.
 
To further determine if the recessive $c$ allele can fix, we must also consider Equation (53) in \cite{Olejarz_2015}, which is the condition, for any number of matings, $n$, for stability of a recessive mutation in workers that causes worker sterility.  Setting $r_1=r^*$ in Equation (53) in \cite{Olejarz_2015}, we obtain
\begin{equation}
\left[ \frac{r^*}{r_\frac{n-1}{n}} - \frac{n\left(1-p_\frac{n-1}{n}\right)}{2} \right] \left[ 2\left(\frac{r^*}{r_\frac{1}{2}}\right) - 1 \right] > 1
\label{eqn:sterility_rec_ster_sta}
\end{equation}
In Condition \eqref{eqn:sterility_rec_ster_sta}, $r_{1/2}$ is the colony reproductive efficiency when a fraction $1/2$ of workers are sterile, $r_{(n-1)/n}$ is the colony reproductive efficiency when a fraction $(n-1)/n$ of workers are sterile, and $p_{(n-1)/n}$ is the fraction of queen-derived males when a fraction $(n-1)/n$ of workers are sterile.  If Condition \eqref{eqn:sterility_rec_ster_sta} is satisfied, then a subsequent recessive mutation, $c$, that acts in workers to render them sterile is evolutionarily stable.

Notice that Condition \eqref{eqn:sterility_rec_ster_inv} depends on the parameters $r_{1/(2n)}$ and $p_{1/(2n)}$, which are related to the effects of the $c$ allele for worker sterility.  Also, notice that Condition \eqref{eqn:sterility_rec_ster_sta} depends on the parameters $r_{1/2}$, $r_{(n-1)/n}$, and $p_{(n-1)/n}$, which are related to the effects of the $c$ allele for worker sterility.  The properties of the particular recessive $c$ allele for worker sterility that is under consideration are therefore essential for determining if the $a$ allele for partial queen control can facilitate the evolution of complete worker sterility.

To gain insight, regarding the parameters $r_{1/(2n)}$, $p_{1/(2n)}$, $r_{1/2}$, $r_{(n-1)/n}$, and $p_{(n-1)/n}$ in Conditions \eqref{eqn:sterility_rec_ster_inv} and \eqref{eqn:sterility_rec_ster_sta}, we can consider the following simple case:
\begin{equation}
\begin{aligned}
p_z &= p' + (1-p') z \\
r_z &= r' + (r^*-r') z \\
\end{aligned}
\label{eqn:parameters_2}
\end{equation}
For the parameter choices given by Equations \eqref{eqn:parameters_2}, Condition \eqref{eqn:sterility_rec_ster_inv} becomes
\begin{equation}
\frac{r^*}{r'} > \frac{3+4n+p'}{3+p'+2n(1+p')}
\label{eqn:sterility_rec_ster_inv_particular}
\end{equation}
Also for the parameter choices given by Equations \eqref{eqn:parameters_2}, Condition \eqref{eqn:sterility_rec_ster_sta} becomes
\begin{equation}
\frac{r^*}{r'} > \frac{4(1-p')+n(3+p')+\sqrt{4(3-p')^2+n^2(3+p')^2+4n(7+p'^2)}}{10+2n+6(n-1)p'}
\label{eqn:sterility_rec_ster_sta_particular}
\end{equation}

To determine if partial queen control can enable the evolutionarily success of a dominant $c$ mutation in workers that renders them sterile, we turn to Equation (20) in \cite{Olejarz_2015}, which is the condition, for any number of matings, $n$, for invasion of a dominant mutation in workers that results in worker sterility:  Setting $r_0=r'$ in Equation (20) in \cite{Olejarz_2015}, this condition becomes
\begin{equation}
\frac{r_\frac{1}{2}}{r'} \left[ 1 + p_\frac{1}{2}\left(\frac{r_\frac{1}{n}}{r'}\right) \right] > 2
\label{eqn:sterility_dom_ster_inv}
\end{equation}
In Condition \eqref{eqn:sterility_dom_ster_inv}, $r_{1/n}$ is the colony reproductive efficiency when a fraction $1/n$ of workers are sterile, $r_{1/2}$ is the colony reproductive efficiency when a fraction $1/2$ of workers are sterile, and $p_{1/2}$ is the fraction of queen-derived males when a fraction $1/2$ of workers are sterile.  If Condition \eqref{eqn:sterility_dom_ster_inv} is satisfied, then a subsequent dominant mutation, $c$, that acts in workers to render them sterile invades a partially queen-controlled population.
 
To further determine if the dominant $c$ allele can fix, we must also consider the equation directly after Equation (49) in \cite{Olejarz_2015}, which is the condition, for any number of matings, $n$, for stability of a dominant mutation in workers that causes worker sterility.  Setting $r_1=r^*$ in the equation directly after Equation (49) in \cite{Olejarz_2015}, we obtain
\begin{equation}
\frac{r^*}{r_\frac{2n-1}{2n}} > \frac{2+3n-np_\frac{2n-1}{2n}}{2(n+1)}
\label{eqn:sterility_dom_ster_sta}
\end{equation}
In Condition \eqref{eqn:sterility_dom_ster_sta}, $r_{(2n-1)/(2n)}$ is the colony reproductive efficiency when a fraction $(2n-1)/(2n)$ of workers are sterile, and $p_{(2n-1)/(2n)}$ is the fraction of queen-derived males when a fraction $(2n-1)/(2n)$ of workers are sterile.  If Condition \eqref{eqn:sterility_dom_ster_sta} is satisfied, then a subsequent dominant mutation, $c$, that acts in workers to render them sterile is evolutionarily stable.

Notice that Condition \eqref{eqn:sterility_dom_ster_inv} depends on the parameters $r_{1/n}$, $r_{1/2}$, and $p_{1/2}$, which are related to the effects of the $c$ allele for worker sterility.  Also, notice that Condition \eqref{eqn:sterility_dom_ster_sta} depends on the parameters $r_{(2n-1)/(2n)}$ and $p_{(2n-1)/(2n)}$, which are related to the effects of the $c$ allele for worker sterility.  The properties of the particular dominant $c$ allele for worker sterility that is under consideration are therefore essential for determining if the $a$ allele for partial queen control can facilitate the evolution of complete worker sterility.

To gain insight, regarding the parameters $r_{1/n}$, $r_{1/2}$, $p_{1/2}$, $r_{(2n-1)/(2n)}$, and $p_{(2n-1)/(2n)}$ in Conditions \eqref{eqn:sterility_dom_ster_inv} and \eqref{eqn:sterility_dom_ster_sta}, we can again consider the simple case given by Equations \eqref{eqn:parameters_2}.  For the parameter choices given by Equations \eqref{eqn:parameters_2}, Condition \eqref{eqn:sterility_dom_ster_inv} becomes
\begin{equation}
\frac{r^*}{r'} > \frac{\sqrt{4n(5-p')(1+p')+4(1+p')^2+n^2(3+p')^2}-n(3+p')}{2(1+p')}
\label{eqn:sterility_dom_ster_inv_particular}
\end{equation}
Also for the parameter choices given by Equations \eqref{eqn:parameters_2}, Condition \eqref{eqn:sterility_dom_ster_sta} becomes
\begin{equation}
\frac{r^*}{r'} > \frac{5+4n-p'}{5-p'+2n(1+p')}
\label{eqn:sterility_dom_ster_sta_particular}
\end{equation}

\begin{figure}
\centering
\begin{subfigure}{0.45\textwidth}
\centering
\caption{}
\includegraphics*[width=1\textwidth]{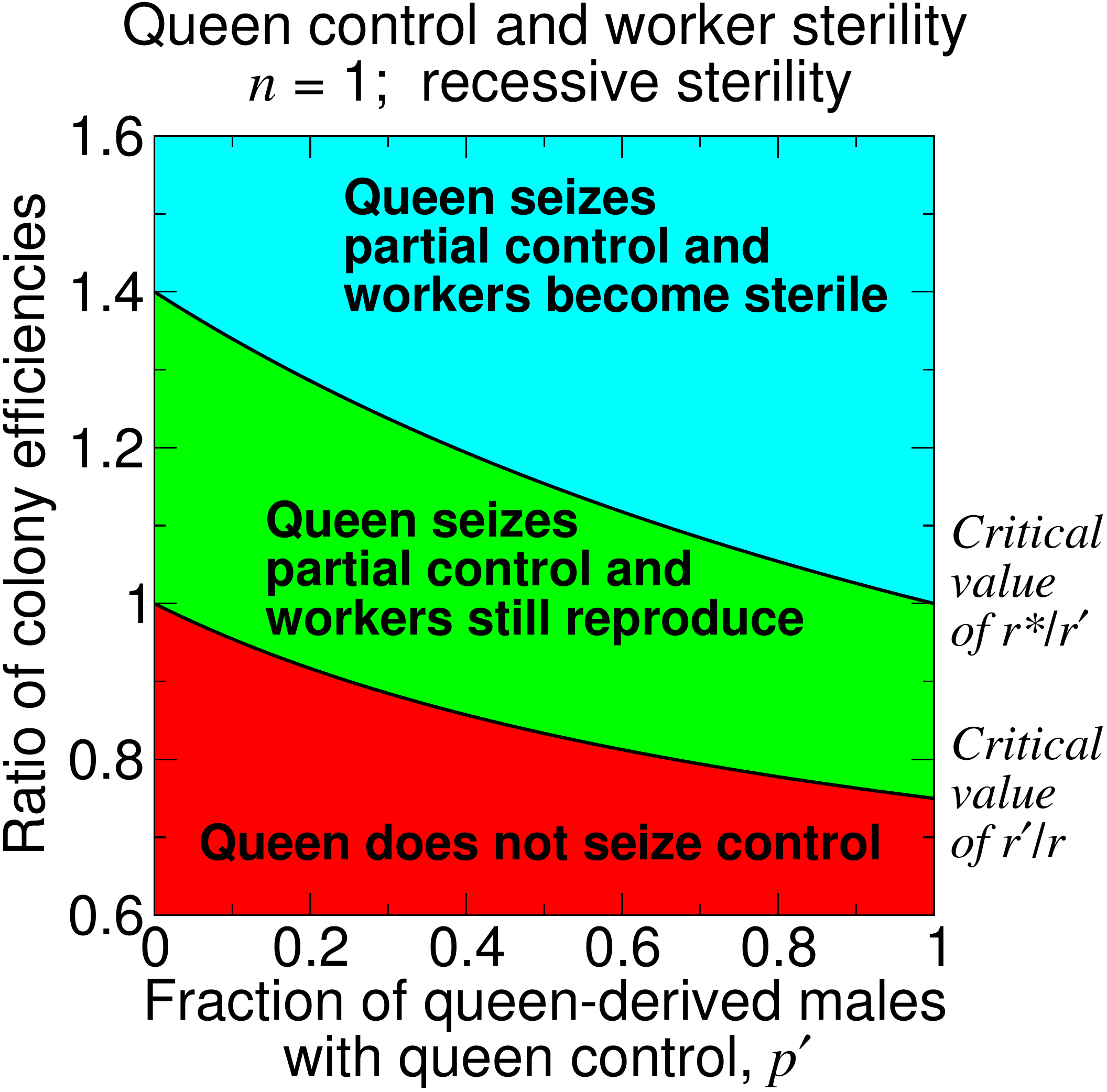}
\end{subfigure}
\quad
\begin{subfigure}{0.45\textwidth}
\centering
\caption{}
\includegraphics*[width=1\textwidth]{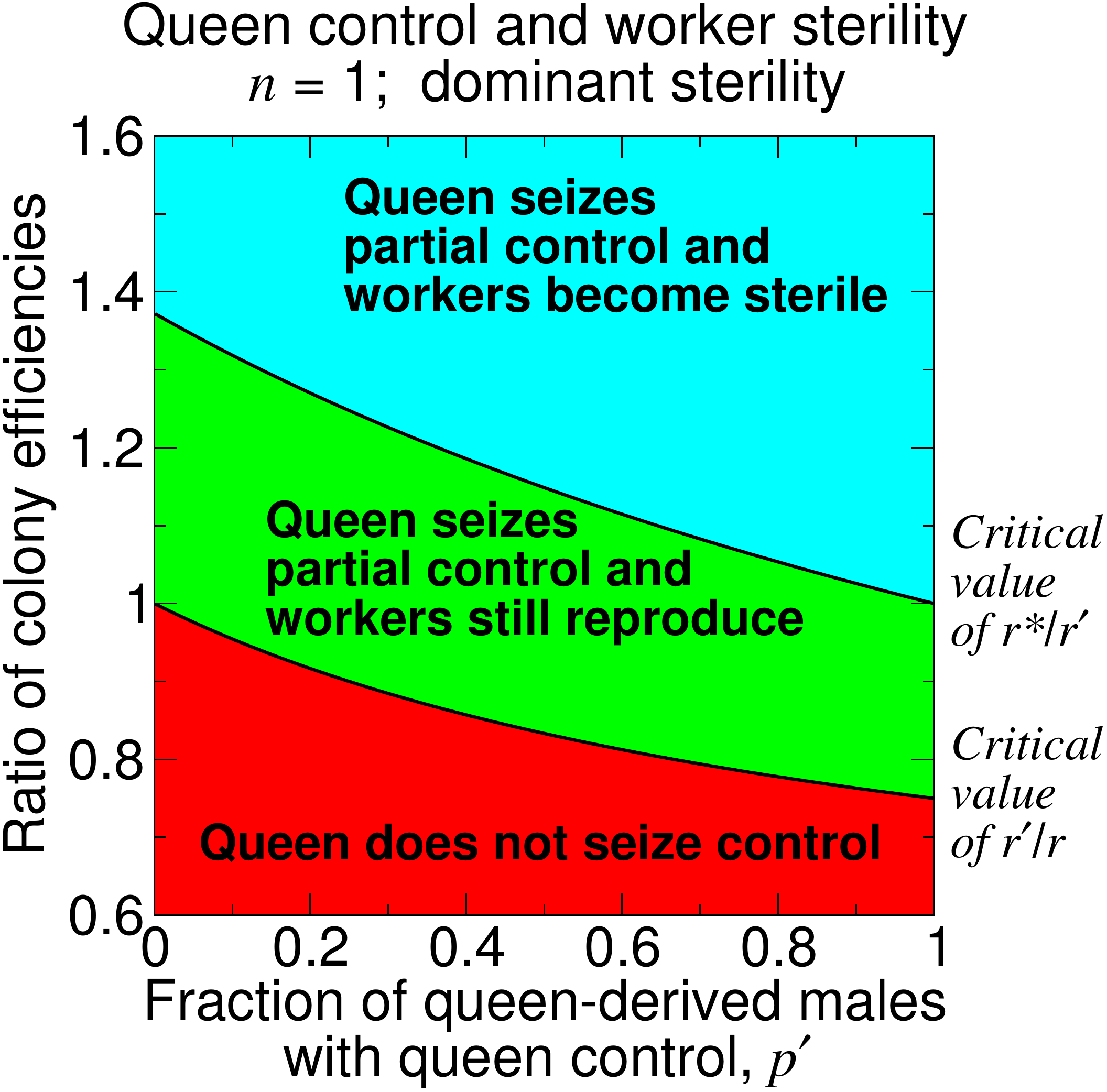}
\end{subfigure}
\par\medskip
\begin{subfigure}{0.45\textwidth}
\centering
\caption{}
\includegraphics*[width=1\textwidth]{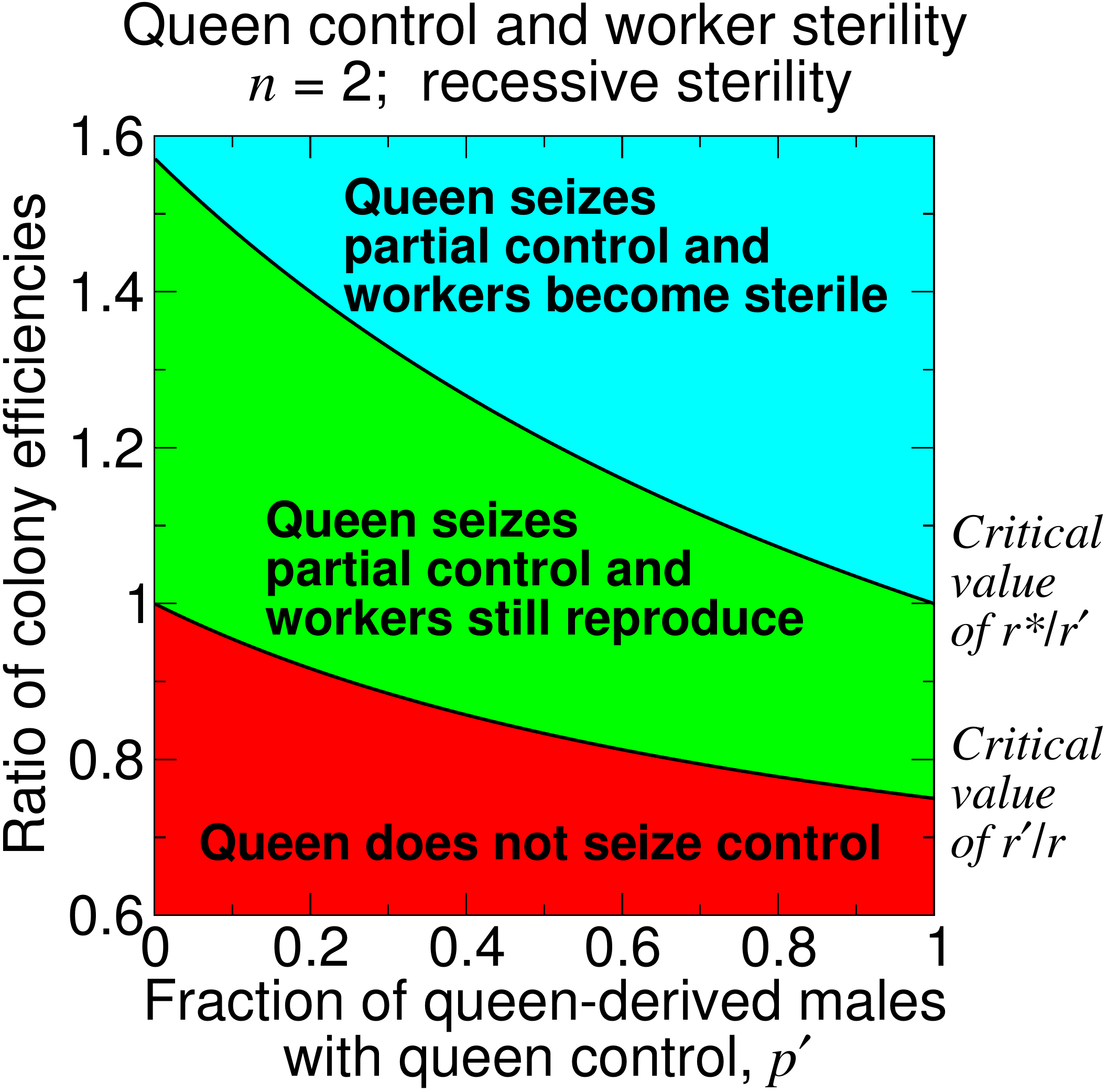}
\end{subfigure}
\quad
\begin{subfigure}{0.45\textwidth}
\centering
\caption{}
\includegraphics*[width=1\textwidth]{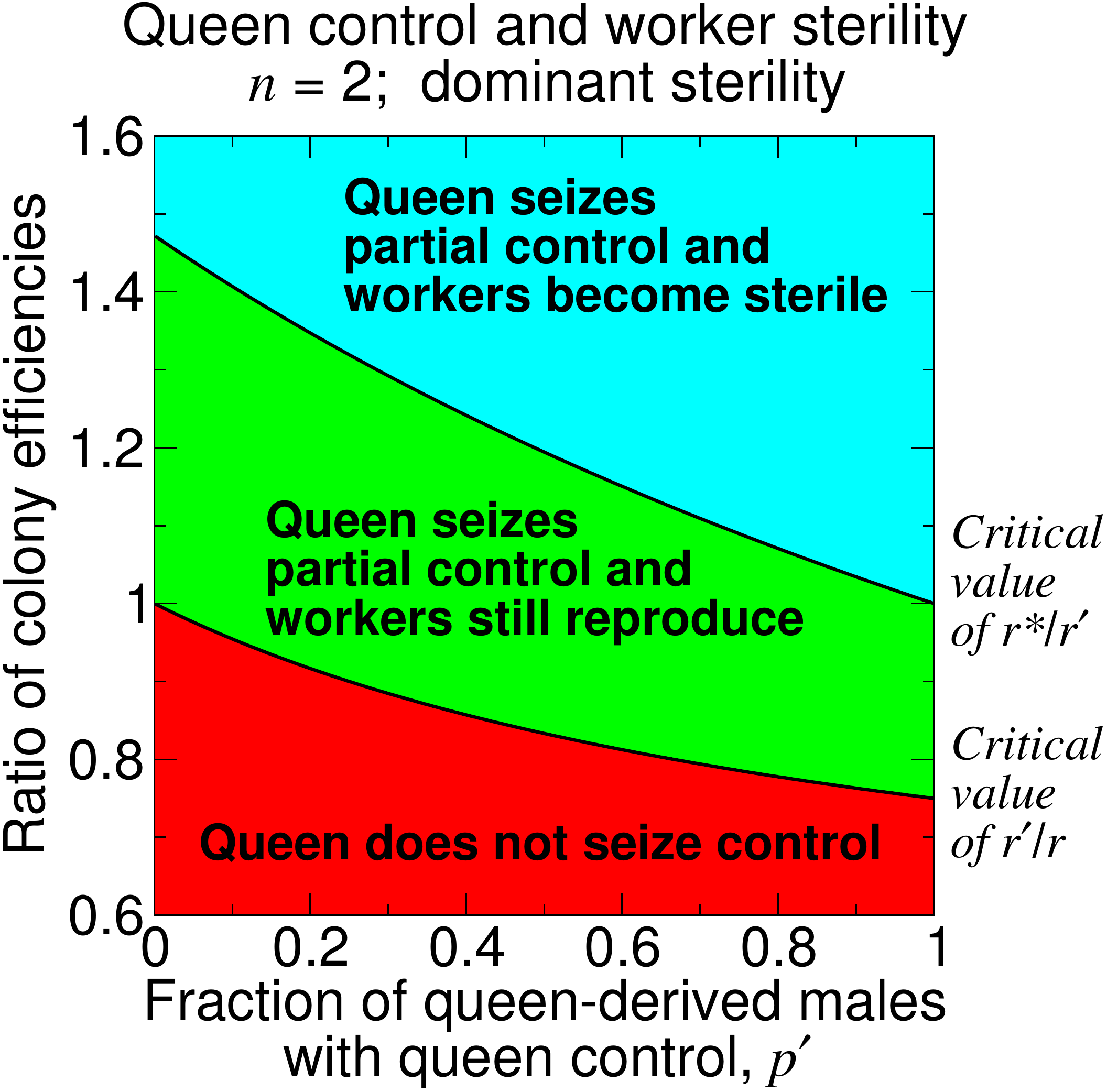}
\end{subfigure}
\caption{A mutation for queen control may or may not induce worker sterility.  Initially, assume that workers are responsible for all male production ($p=0$).  A mutation in queens then causes them to seize partial control of male production ($0<p'<1$).  More powerful queen control (i.e., mutations causing larger values of $p'$) can evolve more easily, since the critical value of $r'/r$ decreases with $p'$.  But more powerful queen control also lowers the critical value of $r^*/r'$ for a subsequent mutation, acting in workers, to render them sterile.  The lower boundary is given by Equation \eqref{eqn:stability}, and the upper boundary is given by (a) Equation \eqref{eqn:sterility_rec_ster_inv_particular} for $n=1$, (b) Equation \eqref{eqn:sterility_dom_ster_inv_particular} for $n=1$, (c) Equation \eqref{eqn:sterility_rec_ster_inv_particular} for $n=2$, and (d) Equation \eqref{eqn:sterility_dom_ster_inv_particular} for $n=2$.  For this plot, we use Equations \eqref{eqn:parameters_2}, and we set $p=0$.  (If we considered $p>0$ instead, then, when plotted between $p<p'<1$ on the horizontal axis, this figure would look qualitatively the same, except that the middle, green region would be smaller.)}
\label{fig:sterility}
\end{figure}

Figure \ref{fig:sterility} shows how partial queen control can facilitate complete worker sterility.  In each panel, the boundary between the lower, red region and the middle, green region is given by Condition \eqref{eqn:stability}.  For values $(p',r'/r)$ in the lower, red region, the queen does not seize partial control.  For values $(p',r'/r)$ in the middle, green region, the queen seizes partial control, and the workers may or may not become sterile.  The boundary between the middle, green region and the upper, blue region is given by Condition \eqref{eqn:sterility_rec_ster_inv_particular} for $n=1$ (Figure \ref{fig:sterility}(a)), Condition \eqref{eqn:sterility_dom_ster_inv_particular} for $n=1$ (Figure \ref{fig:sterility}(b)), Condition \eqref{eqn:sterility_rec_ster_inv_particular} for $n=2$ (Figure \ref{fig:sterility}(c)), and Condition \eqref{eqn:sterility_dom_ster_inv_particular} for $n=2$ (Figure \ref{fig:sterility}(d)).  This boundary determines if workers become sterile after the queen has seized partial control of male production.  Suppose that the queen seizes partial control of male production.  For values $(p',r^*/r')$ in the middle, green region, the $c$ mutation for worker sterility does not invade.  For values $(p',r^*/r')$ in the upper, blue region, the $c$ mutation for worker sterility invades and is evolutionarily stable, rendering workers totally non-reproductive.

\begin{figure}
\centering
\begin{subfigure}{0.45\textwidth}
\centering
\caption{}
\includegraphics*[width=1\textwidth]{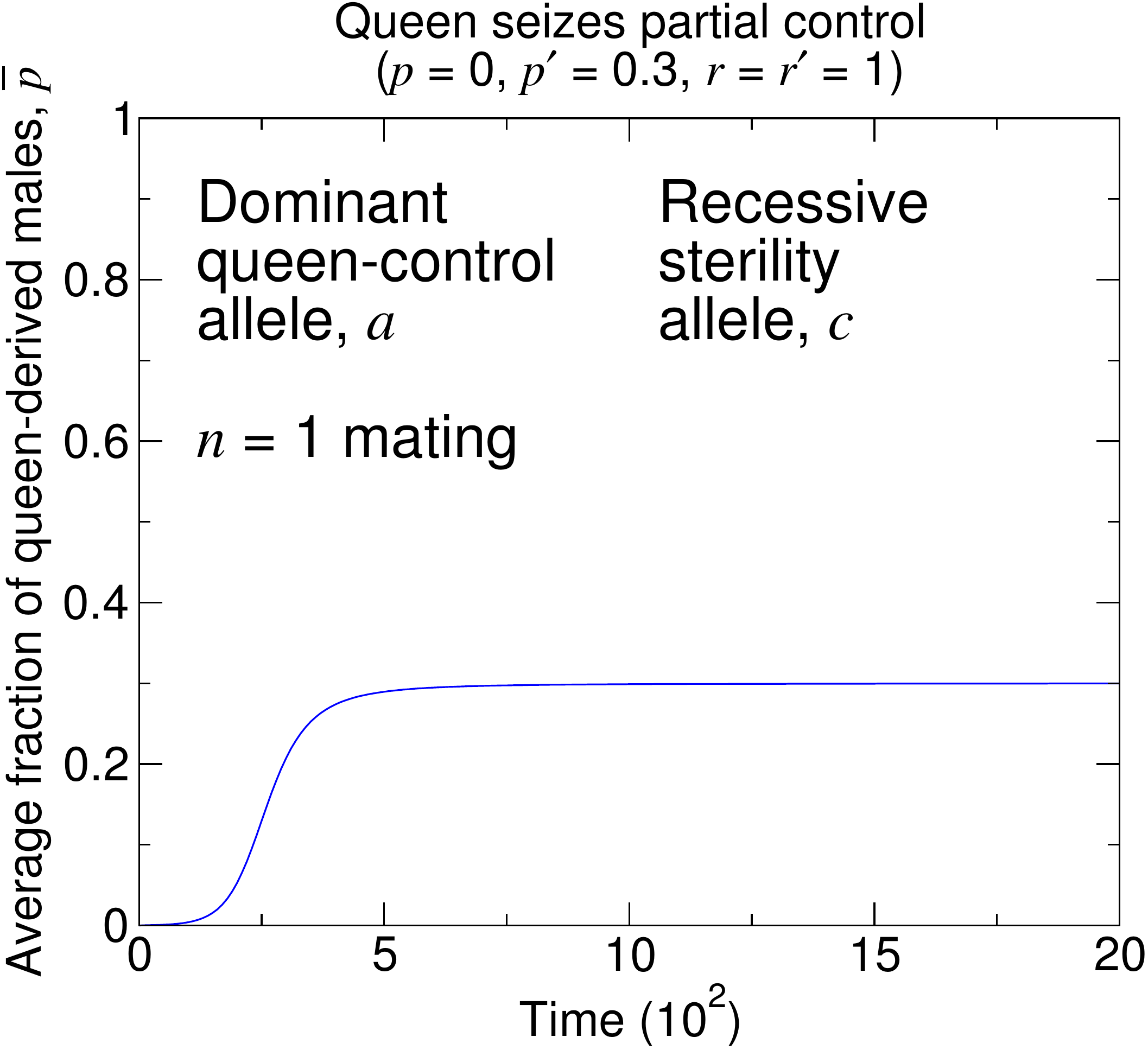}
\end{subfigure}
\quad
\begin{subfigure}{0.45\textwidth}
\centering
\caption{}
\includegraphics*[width=1\textwidth]{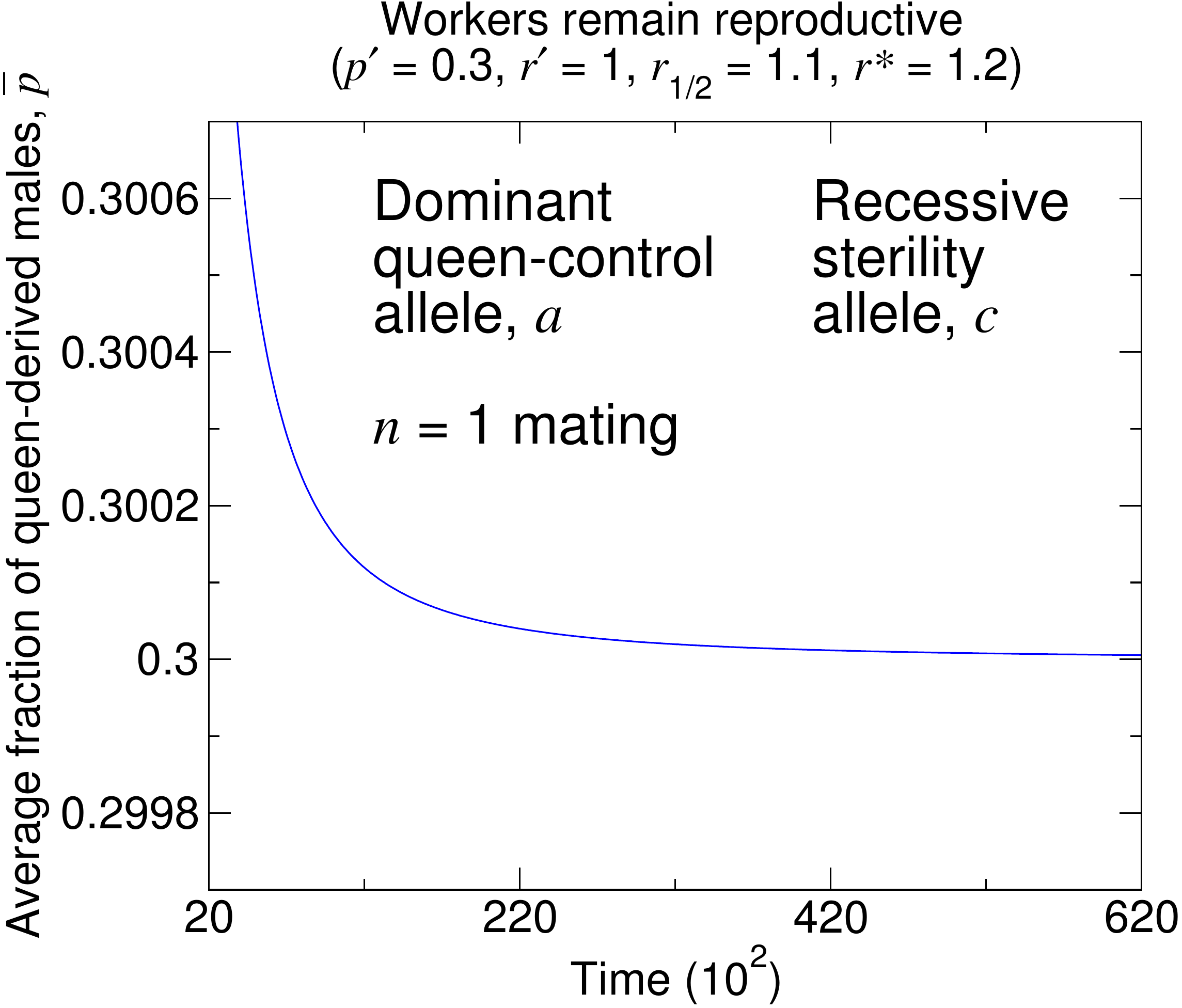}
\end{subfigure}
\par\medskip
\begin{subfigure}{0.45\textwidth}
\centering
\caption{}
\includegraphics*[width=1\textwidth]{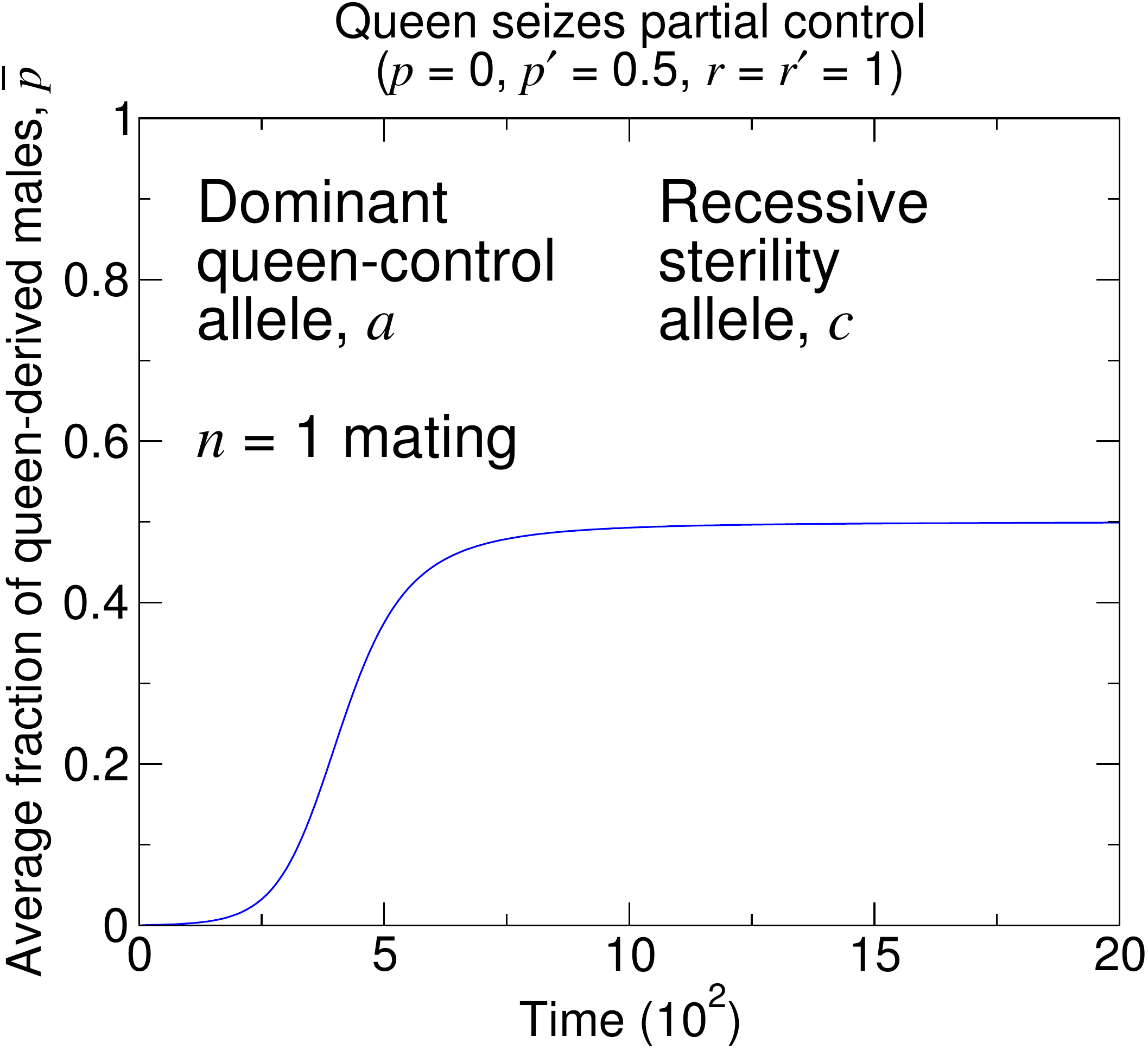}
\end{subfigure}
\quad
\begin{subfigure}{0.45\textwidth}
\centering
\caption{}
\includegraphics*[width=1\textwidth]{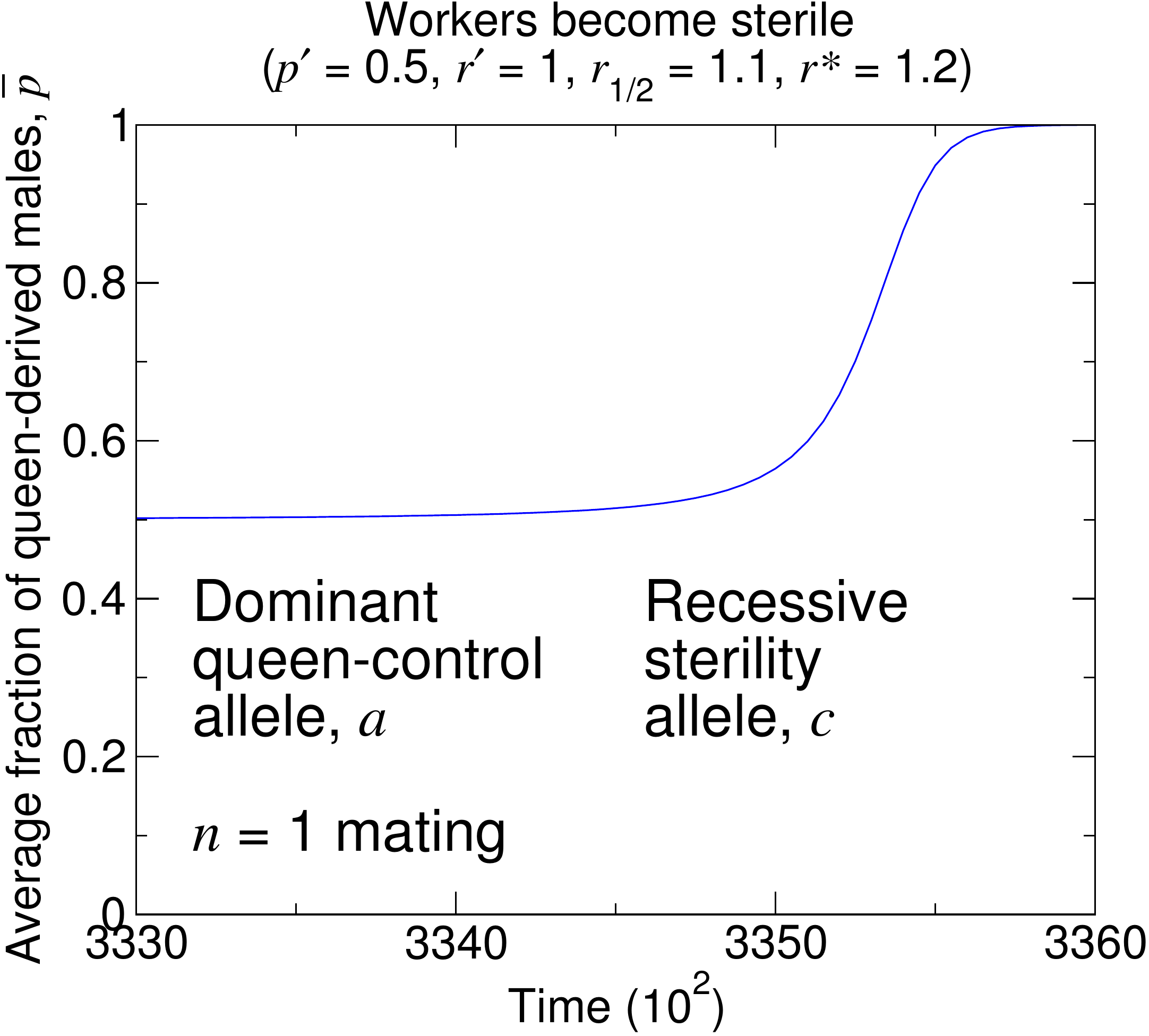}
\end{subfigure}
\caption{Simulations reveal the behaviors shown in Figure \ref{fig:sterility}.  If queens seize a small amount of control over male production (a), then a subsequent mutation, acting in workers, does not cause them to become sterile (b).  If queens seize a large amount of control over male production (c), then a subsequent mutation, acting in workers, causes them to become sterile (d).  Thus, queen control can facilitate the formation of a sterile worker caste.  (In (b) and (d), $r_{1/2}$ denotes the colony efficiency when $1/2$ of workers in the colony have the phenotype for worker sterility.  We follow the assumption for $r_z$ in Equations \eqref{eqn:parameters_2} for determining the value of $r_{1/2}=1.1$ for these simulations.  The initial conditions are (a) $X_{AA,0}=1-10^{-3}$ and $X_{AA,1}=10^{-3}$, (b) $X_{CC,0}=1-10^{-1}$ and $X_{CC,1}=10^{-1}$, (c) $X_{AA,0}=1-10^{-3}$ and $X_{AA,1}=10^{-3}$, and (d) $X_{CC,0}=1-10^{-3}$ and $X_{CC,1}=10^{-3}$.  For (b) and (d), we introduce the $c$ allele for worker sterility at time $t=2000$.)}
\label{fig:sterility_sim}
\end{figure}

Corresponding simulations of the evolutionary dynamics are shown in Figure \ref{fig:sterility_sim}.  The average fraction of queen-derived males in the population, $\overline{p}$, is calculated in the same way as for Figure \ref{fig:rep_sim}.

\section{Discussion}

We have studied, in a haplodiploid population-genetic model of a social Hymenopteran, the conditions for invasion and fixation of genes that act in queens to suppress worker reproduction. We have also studied the conditions under which selection subsequently favors genes that act in workers to resist queen control.

The condition for evolutionary invasion and stability of queen control, Condition \eqref{eqn:stability}, is always easier to satisfy than the conditions for subsequent worker acquiescence; the former condition does not require colony efficiency gains to queen control, while the latter conditions do. Therefore, there always exist regions of parameter space where queen control can invade and fix, but where worker suppression of queen control is subsequently selected for. In these cases, queen control can be thought of as coercive (that is, against workers' evolutionary interests). There also exist regions of parameter space---where queen control invades and fixes, and where the conditions for worker acquiescence are satisfied---where evolved queen control can be thought of as honest signalling (that is, in workers' evolutionary interests). 

We have thus shown that, within the same simple setup, both coercive control and control via honest signalling are possible. This theoretical result is interesting in light of the continuing empirical debate over whether queen control represents coercion or signalling. Many recent works have expressed disfavor toward the coercion hypothesis \citep{Keller_1993,Holman_2010,vanZweden_2013,Chapuisat_2014,Oi_2015b,Peso_2015}, but our results demonstrate that coercive control could have evolved often in the social Hymenoptera. 

The crucial consideration in our analysis is how the establishment of queen control changes the colony's overall reproductive efficiency.  The efficiency increase, $r'/r$, needed for a queen-control allele to be stable to counteraction by workers, given by Conditions \eqref{eqn:resistance_rec_ster_sta_particular} or \eqref{eqn:resistance_dom_ster_sta_particular}, increases with the strength of queen control (i.e., the amount by which $p'=1$ exceeds $p$).  But the efficiency increase, $r^*/r'$, needed for a subsequent allele, acting in workers, to induce their sterility, given by Conditions \eqref{eqn:sterility_rec_ster_inv_particular} or \eqref{eqn:sterility_dom_ster_inv_particular}, decreases with the strength of queen control (i.e., the magnitude of $p'$).  Thus, stronger queen control is more susceptible to worker resistance, but it also more easily selects for worker non-reproduction.  An understanding of the long-term evolutionary consequences of queen control must consider the specific types of mutations that act in workers and incorporate both of these effects.

In our analysis, colony efficiencies with and without queen control were treated as static parameters. However, because queen control directly limits the workers' contribution to the production of drones, it makes it beneficial for workers instead to invest their resources in colony maintenance tasks \citep{Wenseleers_2004,Wenseleers_2006}. Therefore, colony efficiency could change if the evolution of queen-induced worker sterility is followed by the evolution of more efficient helping by workers \citep{Gonzalez_2014,Gonzalez_2015}. Under this scenario, it is possible that queen control establishes in a system where worker resistance is initially under positive selection---Conditions \eqref{eqn:resistance_rec_ster_sta_particular} and \eqref{eqn:resistance_dom_ster_sta_particular} do not hold---but that subsequent efficiency gains by the now-sterile worker caste increase $r'$ sufficiently that Conditions \eqref{eqn:resistance_rec_ster_sta_particular} and \eqref{eqn:resistance_dom_ster_sta_particular} come to hold, so that worker resistance is no longer selected for. 

Our results facilitate a crucial connection with ongoing experimental efforts in sociobiology.  Research is underway on the chemical characteristics of queen-emitted pheromones that induce specific primer or releaser effects on workers \citep{Wagner_1998,Eliyahu_2011,Smith_2012,VanOystaeyen_2014,Bello_2015,Sharma_2015,Yew_2015,Zhou_2015}, and on the molecular mechanisms and gene networks behind reproductive regulation \citep{Thompson_2007,Khila_2008,Khila_2010,Kocher_2010,Fischman_2011,Mullen_2014,Toth_2014,Rehan_2014,Rehan_2015}.  Such experimental programs, together with measurements of the effects of queen control on colony parameters and the mathematical conditions herein, could promote understanding of the precise evolutionary steps that have led to reproductive division of labor.

\vspace*{10mm}

\section*{Acknowledgements}

This work was supported by the John Templeton Foundation and in part by a grant from B. Wu and Eric Larson.

\newpage

\clearpage

\appendix

\section{Supporting Information}
\label{SI:stability}

In this Supporting Information, we derive evolutionary invasion and stability conditions 
for queen control of worker reproduction.  The mathematical structure of our model is 
identical to the model featured in \cite{Olejarz_2016}.  There are three types of females: $AA$, $Aa$, 
and $aa$.  There are two types of males: $A$ and $a$.  The three types of unfertilized 
females are denoted by $x_{AA}$, $x_{Aa}$, and $x_{aa}$.  The two types of males are 
denoted by $y_A$ and $y_a$.  Each colony is headed by a single, reproductive female.  
The $3(n+1)$ types of colonies are denoted by $AA,m$; $Aa,m$; and $aa,m$, where 
$m$ is the number of the queen's matings that were with type $a$ males.  
We naturally have that $0 \leq m \leq n$.  
The mating events are shown in Figure \ref{fig:colonies}(a).  
The reproduction events are shown in Figure \ref{fig:colonies}(b).

The evolutionary dynamics are specified by the following equations:
\begin{equation}
\begin{aligned}
\dot{X}_{AA,m} = \frac{dX_{AA,m}}{dt} &= {n \choose m} x_{AA}y_A^{n-m}y_a^m - \phi X_{AA,m} \\
\dot{X}_{Aa,m} = \frac{dX_{Aa,m}}{dt} &= {n \choose m} x_{Aa}y_A^{n-m}y_a^m - \phi X_{Aa,m} \\
\dot{X}_{aa,m} = \frac{dX_{aa,m}}{dt} &= {n \choose m} x_{aa}y_A^{n-m}y_a^m - \phi X_{aa,m}
\end{aligned}
\label{eqn:SI_dynamics}
\end{equation}
We impose the following density constraint:
\begin{equation}
\sum_{m=0}^n(X_{AA,m}+X_{Aa,m}+X_{aa,m})=1
\label{eqn:SI_density}
\end{equation}
To enforce Equation \eqref{eqn:SI_density}, we set
\begin{equation}
\phi=(x_{AA}+x_{Aa}+x_{aa})(y_A+y_a)^n
\label{eqn:SI_phi}
\end{equation}

\vspace*{10mm}

\subsection{Invasion of a Dominant Queen-Control Allele}
\label{sec:SI_dom_invasion}

We focus on the evolution of the colony frequencies.  Using 
Figure \ref{fig:colonies}(b), we write each type of reproductive of a 
colony ($x_{AA}$, $x_{Aa}$, $x_{aa}$, $y_A$, and $y_a$) as follows:
\begin{equation}
\begin{aligned}
x_{AA} =& \sum_{m=0}^{n} \left[ \frac{n-m}{n}rX_{AA,m} + \frac{n-m}{2n}r'X_{Aa,m} \right] \\
x_{Aa} =& \sum_{m=0}^{n} \left[ \frac{m}{n}rX_{AA,m} + \frac{1}{2}r'X_{Aa,m} + \frac{n-m}{n}r'X_{aa,m} \right] \\
x_{aa} =& \sum_{m=0}^{n} \left[ \frac{m}{2n}r'X_{Aa,m} + \frac{m}{n}r'X_{aa,m} \right] \\
y_{A}  =& \sum_{m=0}^{n} \left[ \frac{2n-m(1-p)}{2n}rX_{AA,m} + \frac{3n-2m+(2m-n)p'}{4n}r'X_{Aa,m} \right. \\
        & \left. + \frac{(n-m)(1-p')}{2n}r'X_{aa,m} \right] \\
y_{a}  =& \sum_{m=0}^{n} \left[ \frac{m(1-p)}{2n}rX_{AA,m} + \frac{2m+n+(n-2m)p'}{4n}r'X_{Aa,m} \right. \\
        & \left. + \frac{n+m+(n-m)p'}{2n}r'X_{aa,m} \right] \\
\end{aligned}
\label{eqn:SI_dom_r_steady_state}
\end{equation}
Among Equations \eqref{eqn:SI_dynamics}, the three equations that are relevant 
for considering invasion of a rare, dominant $a$ allele are
\begin{equation}
\begin{aligned}
\dot{X}_{AA,0} &= x_{AA}y_A^n - \phi X_{AA,0} \\
\dot{X}_{AA,1} &= n x_{AA}y_A^{n-1}y_a - \phi X_{AA,1} \\
\dot{X}_{Aa,0} &= x_{Aa}y_A^n - \phi X_{Aa,0}
\end{aligned}
\label{eqn:SI_dom_dynamics}
\end{equation}
If a small amount of the $a$ allele is introduced into the population, then shortly after 
the perturbation, the colony frequencies have the following form (with $\epsilon \ll 1$):
\begin{equation}
\begin{aligned}
X_{AA,0} &= 1 & -\epsilon\delta^{(1)}_{AA,0} & -\mathcal{O}(\epsilon^2) \\[0.1cm]
X_{AA,1} &=   & +\epsilon\delta^{(1)}_{AA,1} & +\mathcal{O}(\epsilon^2) \\[0.1cm]
X_{Aa,0} &=   & +\epsilon\delta^{(1)}_{Aa,0} & +\mathcal{O}(\epsilon^2)
\end{aligned}
\label{eqn:SI_dom_epsilon}
\end{equation}
Using Equations \eqref{eqn:SI_dom_epsilon}, the density constraint, 
Equation \eqref{eqn:SI_density}, takes the following form:
$\mathcal{O}(\epsilon)$:
\begin{equation}
\delta^{(1)}_{AA,0}=\delta^{(1)}_{AA,1}+\delta^{(1)}_{Aa,0}
\label{eqn:SI_density_1}
\end{equation}

We substitute Equations \eqref{eqn:SI_phi}, \eqref{eqn:SI_dom_r_steady_state}, \eqref{eqn:SI_dom_epsilon}, and \eqref{eqn:SI_density_1} into Equations \eqref{eqn:SI_dom_dynamics}.  We find that the condition for evolution of queen control is that the dominant eigenvalue of the Jacobian matrix in the following equation is greater than zero:
\begin{equation*}
\begin{pmatrix} \dot{\delta}^{(1)}_{AA,1} \\ \dot{\delta}^{(1)}_{Aa,0} \end{pmatrix} = \begin{pmatrix} \frac{-(1+p)r}{2} & \frac{(1+p')nr'}{4} \\ \frac{r}{n} & \frac{r'-2r}{2} \end{pmatrix} \begin{pmatrix} \delta^{(1)}_{AA,1} \\ \delta^{(1)}_{Aa,0} \end{pmatrix}
\end{equation*}
The dominant allele for queen control of worker reproduction increases in frequency if
\begin{equation}
\frac{r'}{r} > \frac{2+2p}{2+p+p'}
\label{eqn:SI_dom_r}
\end{equation}

\vspace*{10mm}

\subsubsection{Alternative derivation}

In an alternative treatment, we can consider evolution in discrete time.  Consider a 
small amount of the mutant allele, $a$, in the population.  $x_{Aa}(T)$ denotes the 
abundance of heterozygous mutant females in generation $T$, and $y_a(T)$ denotes the 
abundance of mutant males in generation $T$.  Assuming that each new generation consists 
only of offspring from the previous generation, what are the abundances of $x_{Aa}$ and 
$y_a$ in the next generation, $T+1$?

Consider the following reproduction events.  $Aa$ females mate with wild-type $A$ males 
at rate $1$ to make $Aa,0$ colonies, and $Aa,0$ colonies make new $Aa$ females at rate $r'/2$.  
$a$ males mate with wild-type $AA$ females at rate $n$ to make $AA,1$ colonies, and $AA,1$ 
colonies make new $Aa$ females at rate $r/n$.  $Aa$ females mate with wild-type $A$ males 
at rate $1$ to make $Aa,0$ colonies, and $Aa,0$ colonies make new $a$ males at rate 
$p'r'/2+(1-p')r'/4=(1+p')r'/4$.  $a$ males mate with wild-type $AA$ females at rate $n$ to 
make $AA,1$ colonies, and $AA,1$ colonies make new $a$ males at rate $(1-p)r/(2n)$.  These 
reproduction events can be summarized as:
\begin{equation}
\begin{pmatrix} x_{Aa}(T+1) \\ y_a(T+1) \end{pmatrix} = \begin{pmatrix} \frac{r'}{2} & r \\ \frac{(1+p')r'}{4} & \frac{(1-p)r}{2} \end{pmatrix} \begin{pmatrix} x_{Aa}(T) \\ y_a(T) \end{pmatrix}
\label{eqn:discrete_matrix}
\end{equation}

The condition for invasion of queen control is that the dominant eigenvalue of the matrix 
in Equation \eqref{eqn:discrete_matrix} is greater than $r$.  Performing this calculation 
gives us Equation \eqref{eqn:SI_dom_r}.

\vspace*{10mm}

\subsection{Invasion of a Recessive Queen-Control Allele}
\label{sec:SI_rec_invasion}

We again focus on evolution of the colony frequencies.  Using Figure \ref{fig:colonies}(b), 
we write each type of reproductive of a colony ($x_{AA}$, $x_{Aa}$, $x_{aa}$, $y_A$, and 
$y_a$) as follows:
\begin{equation}
\begin{aligned}
x_{AA} =& \sum_{m=0}^{n} \left[ \frac{n-m}{n}rX_{AA,m} + \frac{n-m}{2n}rX_{Aa,m} \right] \\
x_{Aa} =& \sum_{m=0}^{n} \left[ \frac{m}{n}rX_{AA,m} + \frac{1}{2}rX_{Aa,m} + \frac{n-m}{n}r'X_{aa,m} \right] \\
x_{aa} =& \sum_{m=0}^{n} \left[ \frac{m}{2n}rX_{Aa,m} + \frac{m}{n}r'X_{aa,m} \right] \\
y_{A}  =& \sum_{m=0}^{n} \left[ \frac{2n-m(1-p)}{2n}rX_{AA,m} + \frac{3n-2m+(2m-n)p}{4n}rX_{Aa,m} \right. \\
        & \left. + \frac{(n-m)(1-p')}{2n}r'X_{aa,m} \right] \\
y_{a}  =& \sum_{m=0}^{n} \left[ \frac{m(1-p)}{2n}rX_{AA,m} + \frac{2m+n+(n-2m)p}{4n}rX_{Aa,m} \right. \\
       & \left. + \frac{n+m+(n-m)p'}{2n}r'X_{aa,m} \right] \\
\end{aligned}
\label{eqn:SI_rec_r_steady_state}
\end{equation}
Among Equations \eqref{eqn:SI_dynamics}, the six equations that are relevant 
for considering invasion of a rare, recessive $a$ allele are
\begin{equation}
\begin{aligned}
\dot{X}_{AA,0} &= x_{AA}y_A^n - \phi X_{AA,0} \\
\dot{X}_{AA,1} &= n x_{AA}y_A^{n-1}y_a - \phi X_{AA,1} \\
\dot{X}_{Aa,0} &= x_{Aa}y_A^n - \phi X_{Aa,0} \\
\dot{X}_{AA,2} &= \frac{n(n-1)}{2} x_{AA}y_A^{n-2}y_a^2 - \phi X_{AA,2} \\
\dot{X}_{Aa,1} &= n x_{Aa}y_A^{n-1}y_a - \phi X_{Aa,1} \\
\dot{X}_{aa,0} &= x_{aa}y_A^n - \phi X_{aa,0}
\end{aligned}
\label{eqn:SI_rec_dynamics}
\end{equation}
If a small amount of the $a$ allele is introduced into the population, then shortly after 
the perturbation, the colony frequencies have the following form (with $\epsilon \ll 1$):
\begin{equation}
\begin{aligned}
X_{AA,0} &= 1 & -\epsilon\delta^{(1)}_{AA,0} & -\epsilon^2\delta^{(2)}_{AA,0} & -\mathcal{O}(\epsilon^3) \\[0.1cm]
X_{AA,1} &=   & +\epsilon\delta^{(1)}_{AA,1} & +\epsilon^2\delta^{(2)}_{AA,1} & +\mathcal{O}(\epsilon^3) \\[0.1cm]
X_{Aa,0} &=   & +\epsilon\delta^{(1)}_{Aa,0} & +\epsilon^2\delta^{(2)}_{Aa,0} & +\mathcal{O}(\epsilon^3) \\[0.1cm]
X_{AA,2} &=   &                              & +\epsilon^2\delta^{(2)}_{AA,2} & +\mathcal{O}(\epsilon^3) \\[0.1cm]
X_{Aa,1} &=   &                              & +\epsilon^2\delta^{(2)}_{Aa,1} & +\mathcal{O}(\epsilon^3) \\[0.1cm]
X_{aa,0} &=   &                              & +\epsilon^2\delta^{(2)}_{aa,0} & +\mathcal{O}(\epsilon^3)
\end{aligned}
\label{eqn:SI_rec_epsilon}
\end{equation}
Equations \eqref{eqn:SI_rec_epsilon}, together with the density 
constraint, Equation \eqref{eqn:SI_density}, yield 
Equation \eqref{eqn:SI_density_1} at $\mathcal{O}(\epsilon)$.  
At $\mathcal{O}(\epsilon^2)$, the density constraint, 
Equation \eqref{eqn:SI_density}, takes the following form:
\begin{equation}
\delta^{(2)}_{AA,0}=\delta^{(2)}_{AA,1}+\delta^{(2)}_{Aa,0}+\delta^{(2)}_{AA,2}+\delta^{(2)}_{Aa,1}+\delta^{(2)}_{aa,0}
\label{eqn:SI_density_2}
\end{equation}

We substitute Equations \eqref{eqn:SI_phi}, \eqref{eqn:SI_rec_r_steady_state}, \eqref{eqn:SI_rec_epsilon}, and \eqref{eqn:SI_density_1} into Equations \eqref{eqn:SI_rec_dynamics}.  At $\mathcal{O}(\epsilon)$, we have
\begin{equation}
\begin{pmatrix} \dot{\delta}^{(1)}_{AA,1} \\ \dot{\delta}^{(1)}_{Aa,0} \end{pmatrix} = \begin{pmatrix} \frac{-(1+p)r}{2} & \frac{(1+p)nr}{4} \\ \frac{r}{n} & \frac{-r}{2} \end{pmatrix} \begin{pmatrix} \delta^{(1)}_{AA,1} \\ \delta^{(1)}_{Aa,0} \end{pmatrix}
\label{eqn:SI_rec_matrix}
\end{equation}
The dominant eigenvalue of the matrix in \eqref{eqn:SI_rec_matrix} is zero, and the corresponding eigenvector is
\begin{equation}
\begin{pmatrix} \delta^{(1)}_{AA,1} \\ \delta^{(1)}_{Aa,0} \end{pmatrix} = \frac{\delta^{(1)}_{AA,0}}{n+2} \begin{pmatrix} n \\ 2 \end{pmatrix}
\label{eqn:SI_rec_eigenvector}
\end{equation}
We therefore use \eqref{eqn:SI_rec_eigenvector} in the following calculations.

We then substitute Equations \eqref{eqn:SI_phi}, \eqref{eqn:SI_rec_r_steady_state}, \eqref{eqn:SI_rec_epsilon}, \eqref{eqn:SI_rec_eigenvector}, and \eqref{eqn:SI_density_2} into Equations \eqref{eqn:SI_rec_dynamics}.  At $\mathcal{O}(\epsilon^2)$, we have
\begin{equation}
\begin{aligned}
-\dot{\delta}^{(2)}_{AA,0}r^{-(n+1)} =& \; \frac{2-n-np}{4n}\left(-2\delta^{(2)}_{AA,1}+n\delta^{(2)}_{Aa,0}\right) \\
                                      & +\frac{-2+np}{n}\delta^{(2)}_{AA,2} \\
                                      & -\frac{n^2+2+n(n-2)p}{4n}\delta^{(2)}_{Aa,1} \\
                                      & +\frac{2-(2+n+np')r'r^{-1}}{2}\delta^{(2)}_{aa,0} \\
                                      & +\frac{n(n+3)}{2(n+2)^2}\left[\delta^{(1)}_{AA,0}\right]^2
\label{eqn:SI_rec_r_d_dt_AA0}
\end{aligned}
\end{equation}
We also have
\begin{equation*}
\begin{aligned}
 \dot{\delta}^{(2)}_{AA,1}r^{-(n+1)} =& \; \frac{1+p}{4}\left(-2\delta^{(2)}_{AA,1}+n\delta^{(2)}_{Aa,0}\right) \\
                                      & +(1-p)\delta^{(2)}_{AA,2} \\
                                      & +\frac{n+2+(n-2)p}{4}\delta^{(2)}_{Aa,1} \\
                                      & +\frac{(1+p')nr'r^{-1}}{2}\delta^{(2)}_{aa,0} \\
                                      & -\frac{n(n+1)}{(n+2)^2}\left[\delta^{(1)}_{AA,0}\right]^2 \\
 \dot{\delta}^{(2)}_{Aa,0}r^{-(n+1)} =& \; \frac{-1}{2n}\left(-2\delta^{(2)}_{AA,1}+n\delta^{(2)}_{Aa,0}\right) \\
                                      & +\frac{2}{n}\delta^{(2)}_{AA,2} \\
                                      & +\frac{1}{2}\delta^{(2)}_{Aa,1} \\
                                      & +r'r^{-1}\delta^{(2)}_{aa,0} \\
                                      & -\frac{2n}{(n+2)^2}\left[\delta^{(1)}_{AA,0}\right]^2 \\
 \dot{\delta}^{(2)}_{AA,2}r^{-(n+1)} =& \; -\delta^{(2)}_{AA,2}+\frac{n(n-1)}{2(n+2)^2}\left[\delta^{(1)}_{AA,0}\right]^2 \\
 \dot{\delta}^{(2)}_{Aa,1}r^{-(n+1)} =& \; -\delta^{(2)}_{Aa,1}+\frac{2n}{(n+2)^2}\left[\delta^{(1)}_{AA,0}\right]^2 \\
 \dot{\delta}^{(2)}_{aa,0}r^{-(n+1)} =& \; -\delta^{(2)}_{aa,0}+\frac{1}{2n}\delta^{(2)}_{Aa,1}
\end{aligned}
\label{eqn:SI_rec_derivative_AAA_2_r}
\end{equation*}
Integrating the equation for $\dot{\delta}^{(2)}_{AA,2}$, we get
\begin{equation}
\delta^{(2)}_{AA,2} = \frac{n(n-1)}{2(n+2)^2}\left[\delta^{(1)}_{AA,0}\right]^2[1-\exp(-r^{n+1}t)]
\label{eqn:SI_rec_r_AA2}
\end{equation}
Integrating the equation for $\dot{\delta}^{(2)}_{Aa,1}$, we get
\begin{equation}
\delta^{(2)}_{Aa,1} = \frac{2n}{(n+2)^2}\left[\delta^{(1)}_{AA,0}\right]^2[1-\exp(-r^{n+1}t)]
\label{eqn:SI_rec_r_Aa1}
\end{equation}
Using the solution for $\delta^{(2)}_{Aa,1}$ to solve for $\delta^{(2)}_{aa,0}$, we get
\begin{equation}
\delta^{(2)}_{aa,0} = \frac{1}{(n+2)^2}\left[\delta^{(1)}_{AA,0}\right]^2[1-(1+r^{n+1}t)\exp(-r^{n+1}t)]
\label{eqn:SI_rec_r_aa0}
\end{equation}
The equations for $\dot{\delta}^{(2)}_{AA,1}$ and $\dot{\delta}^{(2)}_{Aa,0}$ can be 
manipulated to yield
\begin{equation*}
\begin{aligned}
r^{-(n+1)}\frac{d}{dt}\left(-2\delta^{(2)}_{AA,1}+n\delta^{(2)}_{Aa,0}\right) =& \; \frac{-(2+p)}{2}\left(-2\delta^{(2)}_{AA,1}+n\delta^{(2)}_{Aa,0}\right) \\
                                                                               & +2p\delta^{(2)}_{AA,2} \\
                                                                               & -\frac{2+(n-2)p}{2}\delta^{(2)}_{Aa,1} \\
                                                                               & -np'r'r^{-1}\delta^{(2)}_{aa,0} \\
                                                                               & +\frac{2n}{(n+2)^2}\left[\delta^{(1)}_{AA,0}\right]^2 \\
\end{aligned}
\end{equation*}
Integrating this equation to solve for the quantity 
$-2\delta^{(2)}_{AA,1}+n\delta^{(2)}_{Aa,0}$, we obtain
\begin{equation}
\begin{aligned}
-2\delta^{(2)}_{AA,1}+n\delta^{(2)}_{Aa,0} =& \; \frac{2n(p-p'r'r^{-1})}{(n+2)^2(2+p)}\left[\delta^{(1)}_{AA,0}\right]^2 \\
                                            & +\frac{2n((2-p)(p-p'r'r^{-1})+pp'r'r^nt)}{(n+2)^2p^2}\left[\delta^{(1)}_{AA,0}\right]^2\exp(-r^{n+1}t) \\
                                            & -\frac{8n(p-p'r'r^{-1})}{(n+2)^2p^2(2+p)}\left[\delta^{(1)}_{AA,0}\right]^2\exp\left(\frac{-(2+p)}{2}r^{n+1}t\right)
\end{aligned}
\label{eqn:SI_rec_r_AA1_Aa0}
\end{equation}

The queen-control allele invades a resident wild-type population if
\begin{equation}
\lim_{\substack{\epsilon t \rightarrow 0 \\ t \rightarrow \infty}}\dot{\delta}^{(2)}_{AA,0}>0
\label{eqn:SI_rec_r_limit}
\end{equation}
Substituting \eqref{eqn:SI_rec_r_d_dt_AA0}, \eqref{eqn:SI_rec_r_AA2}, 
\eqref{eqn:SI_rec_r_Aa1}, \eqref{eqn:SI_rec_r_aa0}, and \eqref{eqn:SI_rec_r_AA1_Aa0} 
into \eqref{eqn:SI_rec_r_limit}, we find that the recessive allele for queen control 
of worker reproduction increases in frequency if
\begin{equation}
\frac{r'}{r} > \frac{2+2p}{2+p+p'}
\label{eqn:SI_rec_r}
\end{equation}

\vspace*{10mm}

\subsection{Stability of a Dominant Queen-Control Allele}

Suppose that we initially have a population in which all queens suppress their workers' reproduction.  If we introduce a small amount of the allele for no queen control, $A$, and if the queen-control allele, $a$, is dominant, then is queen control evolutionarily stable to being undone by non-controlling queens?

Based on what we already have, the evolutionary stability condition for a dominant queen-control allele is obtained readily using a simplified procedure.  Notice that the calculations of Section \ref{sec:SI_rec_invasion} for invasion of a recessive queen-control allele describe the following scenario:  We begin with a homogeneous population of colonies, where all individuals are homozygous for the $A$ allele.  A fraction $p$ of males originate from the queen, and each colony's reproductive efficiency is $r$.  The mutant allele's only effects are to change the value of $p'$ relative to $p$ and to alter the colony efficiency $r'$ relative to $r$.  Here, $p$ and $r$ are the fraction of queen-derived males and the colony efficiency, respectively, for colonies headed by type $AA$ and type $Aa$ queens, while $p'$ and $r'$ are the corresponding parameters for colonies headed by type $aa$ queens.

Then consider the evolutionary stability of a dominant $a$ allele for controlling queens.  We again begin with a homogeneous population of colonies, but in this case, all individuals are homozygous for the $a$ allele.  A fraction $p'$ of males originate from the queen, and each colony's reproductive efficiency is $r'$.  The mutant allele's only effects are to change the value of $p$ relative to $p'$ and to alter the colony efficiency $r$ relative to $r'$.  Here, $p'$ and $r'$ are the fraction of queen-derived males and the colony efficiency, respectively, for colonies headed by type $aa$ and type $Aa$ queens, while $p$ and $r$ are the corresponding parameters for colonies headed by type $AA$ queens.

Therefore, if we take Equation \eqref{eqn:SI_rec_r}, swap $p$ and $p'$, swap $r$ and $r'$, and reverse the sign of the inequality, then we obtain the condition for evolutionary stability of a dominant queen-control allele:
\begin{equation*}
\frac{r'}{r} > \frac{2+p+p'}{2+2p'}
\end{equation*}

\vspace*{10mm}

\subsection{Stability of a Recessive Queen-Control Allele}

Suppose that we initially have a population in which all queens suppress their workers' reproduction.  If we introduce a small amount of the allele for no queen control, $A$, and if the queen-control allele, $a$, is recessive, then is queen control evolutionarily stable to being undone by non-controlling queens?

Based on what we already have, the evolutionary stability condition for a recessive queen-control allele is obtained readily using a simplified procedure.  Notice that the calculations of Section \ref{sec:SI_dom_invasion} for invasion of a dominant queen-control allele describe the following scenario:  We begin with a homogeneous population of colonies, where all individuals are homozygous for the $A$ allele.  A fraction $p$ of males originate from the queen, and each colony's reproductive efficiency is $r$.  The mutant allele's only effects are to change the value of $p'$ relative to $p$ and to alter the colony efficiency $r'$ relative to $r$.  Here, $p$ and $r$ are the fraction of queen-derived males and the colony efficiency, respectively, for colonies headed by type $AA$ queens, while $p'$ and $r'$ are the corresponding parameters for colonies headed by type $Aa$ and type $aa$ queens.

Then consider the evolutionary stability of a recessive $a$ allele for controlling queens.  We again begin with a homogeneous population of colonies, but in this case, all individuals are homozygous for the $a$ allele.  A fraction $p'$ of males originate from the queen, and each colony's reproductive efficiency is $r'$.  The mutant allele's only effects are to change the value of $p$ relative to $p'$ and to alter the colony efficiency $r$ relative to $r'$.  Here, $p'$ and $r'$ are the fraction of queen-derived males and the colony efficiency, respectively, for colonies headed by type $aa$ queens, while $p$ and $r$ are the corresponding parameters for colonies headed by type $AA$ and type $Aa$ queens.

Therefore, if we take Equation \eqref{eqn:SI_dom_r}, swap $p$ and $p'$, swap $r$ and $r'$, and reverse the sign of the inequality, then we obtain the condition for evolutionary stability of a recessive queen-control allele:
\begin{equation*}
\frac{r'}{r} > \frac{2+p+p'}{2+2p'}
\end{equation*}

\end{document}